\documentclass[11pt,letterpaper]{article}
\usepackage[margin=1in]{geometry}
\usepackage{microtype}
\usepackage{amsmath,amssymb,amsthm,mathtools}
\usepackage{booktabs,array}
\usepackage{float}
\usepackage{algorithm}
\usepackage[noend]{algpseudocode}
\algrenewcommand{\algorithmicrequire}{\textbf{Input:}}
\usepackage{graphicx}
\usepackage{pgfplots}
\pgfplotsset{compat=1.18}
\usepackage[hidelinks]{hyperref}
\usepackage[nameinlink,capitalize,noabbrev]{cleveref}
\makeatletter
\def\theHALG@line{\thealgorithm.\arabic{ALG@line}}
\makeatother
\newtheorem{theorem}{Theorem}[section]
\newtheorem{lemma}[theorem]{Lemma}
\newtheorem{proposition}[theorem]{Proposition}
\newtheorem{corollary}[theorem]{Corollary}
\theoremstyle{definition}
\newtheorem{definition}[theorem]{Definition}
\theoremstyle{remark}
\newtheorem{remark}[theorem]{Remark}
\newcommand{\MMS}{\operatorname{MMS}}
\newcommand{\Q}{\mathcal Q}
\newcommand{\R}{\mathbb R}

\newcommand{\E}{\mathbb E}
\newcommand{\Prb}{\mathbb P}
\newcommand{\one}{\mathbf 1}

\DeclareMathOperator{\supp}{supp}
\hypersetup{pdftitle={Toward Constant-Factor Maximin Share Guarantees for Subadditive Valuations},pdfauthor={}}
\title{Toward Constant-Factor Maximin Share Guarantees for Subadditive Valuations\thanks{The algorithms and their analyses in this paper were developed through extensive interactions with AI tools, especially GPT 5.6 and GPT 6. These tools were also used to organize the results and revise the presentation. The authors have worked to verify the results and proofs and to make the text accessible to human readers. We take full responsibility for the final text and results, including any remaining errors or omissions.}\thanks{We thank Uriel Feige for his helpful comments on the first draft of this paper, especially on the result for general subadditive valuations.}}
\author{Bo Li\\
  {\small Hong Kong Polytechnic University}\\
  {\small\href{mailto:comp-bo.li@polyu.edu.hk}{\texttt{comp-bo.li@polyu.edu.hk}}}
  \and
  Ankang Sun\\
  {\small Shandong University}\\
  {\small\href{mailto:ankang.sun@sdu.edu.cn}{\texttt{ankang.sun@sdu.edu.cn}}}}
\date{}
\begin{document}

\maketitle
\begin{abstract}
For $n$ agents and $m$ indivisible goods, an agent's maximin share (MMS) is the largest minimum bundle value she can obtain by partitioning the goods into $n$ bundles.  Constant MMS approximations are known for additive, submodular, or XOS valuations, but whether such a guarantee exists for general subadditive valuations remains open. We advance this question in two directions. For general subadditive valuations, we exponentially improve the guarantee from $\Omega(1/\log\log n)$ to $\Omega(1/\log^* n)$, where $\log^* n$ is the iterated logarithm. We then seek constant MMS guarantees under additional structure on subadditive valuations by studying $q$-partitioning valuations: every partition of a bundle into at most $q$ groups admits nonnegative group prices that sum to the bundle's value and whose sum on any collection of groups is at most the value of their union. This hierarchy ranges from subadditivity at $q=2$ to XOS for $q\ge m$. We give the first study of MMS in the intermediate classes and prove constant guarantees whenever $q/n$ has a fixed positive lower bound, including $1/6$ for $q\ge n$ and $1/4$ for $q\ge2n$. For XOS valuations, we further improve the universal guarantee from $4/17\approx 0.2353$ to $5/16=0.3125$.

\medskip

For subadditive valuations, we prove a bidding amplification theorem that gives an exponential bound on each agent's probability of receiving less than half her MMS, while satisfying her budget in every outcome. For $q$-partitioning valuations, we give a scalar criterion that converts bounds on the value lost from disjoint deletions within a bundle into bounds on potential losses, even when an agent contributes zero below the target bundle value. For XOS valuations, we prove a reservation theorem that selects goods for integral assignment together with auxiliary additive valuations and accounts for the loss in rounding.
\end{abstract}

\newpage

\section{Introduction}
We study maximin share (MMS) fair allocation of $m$ indivisible goods among $n$ agents with different preferences. An agent's MMS is the largest value she can guarantee by partitioning the goods into $n$ bundles and receiving a bundle she values least~\cite{Budish11}. This benchmark adapts the familiar divide-and-choose principle to indivisible goods: the agent chooses the partition, but has no control over which bundle she receives. Different agents may choose different partitions.
An allocation that gives every agent at least her MMS may not exist, even when valuations are additive~\cite{PW14}. We therefore ask for approximate MMS allocations, which give every agent at least a specified fraction of her MMS. Early work proved constant-factor guarantees for additive valuations~\cite{PW14}, submodular valuations~\cite{BK17}, and fractionally subadditive (XOS) valuations~\cite{GHS+22}. We review subsequent improvements in Section~\ref{sec:related}.

In this paper, we study the more general class of \emph{subadditive} valuations, for which the value of a union of sets is at most the sum of their values. This class contains all three classes above. Subadditive valuations provide a general model of complement-free preferences and play an important role in welfare maximization, budget-feasible procurement, and online allocation~\cite{DNS10,Feige09,NPS24,MNPS+26}. In fair division, Ghodsi et al.~\cite{GHS+22} first obtained an $\Omega(1/\log m)$ MMS guarantee. Subsequent work replaced the dependence on $m$ with a dependence on $n$~\cite{SS22,FH25} and improved the guarantee to $\Omega(1/\log\log n)$~\cite{SS25,Feige25}. 
Despite this progress, the existence of a constant-factor MMS allocation remains open:
\begin{quote}
Does every instance with subadditive valuations admit an allocation giving every agent at least a universal constant fraction of her maximin share?
\end{quote}

We approach the above question in two directions. First, for arbitrary subadditive valuations, we improve the guarantee from $\Omega(1/\log\log n)$ to $\Omega(1/\log^* n)$, where $\log^* n$ is the iterated logarithm. Second, we ask how much additional structure suffices for a constant guarantee and study the hierarchy introduced by Bangachev and Weinberg~\cite{BW25}, which interpolates between subadditive and XOS valuations. We obtain constant MMS guarantees for intermediate classes that can strictly contain XOS. At the XOS endpoint, we use the stronger item-level additive structure to further improve the universal guarantee from $4/17$ to $5/16$~\cite{FG25}.

\subsection{Main results}
Suppose each agent $i$ has a monotone subadditive valuation $v_i$ and MMS $\mu_i$. An $\alpha$-MMS allocation consists of pairwise disjoint bundles $A_i$ such that $v_i(A_i)\ge\alpha\mu_i$ for every agent. Our paper focuses on the existence of such allocations.

\paragraph{An iterated-logarithmic approximation for subadditive valuations.}
Our first result improves the best general guarantee from $\Omega(1/\log\log n)$~\cite{SS25,Feige25} to $\Omega(1/\log^* n)$.
To state the
precise factor, we define $d(n)=1+\log_2^* n$ and $D(n)=2^{\lceil\log_2 d(n)\rceil}$; that is $D(n)$ is the smallest power of two at least $d(n)$. Let
\[
 \alpha(n)=\frac{1}{4D(n)-2}.
\]

\begin{theorem}[General subadditive valuations]\label{thm:sa-main}
Every instance with $n\ge2$ agents and subadditive valuations admits an $\alpha(n)$-MMS allocation. In particular,
\[
 \alpha(n)>\frac{1}{8(1+\log_2^* n)}.
\]
\end{theorem}

To prove \cref{thm:sa-main}, we construct, for each agent, a bundle worth at least half of her MMS. These bundles may overlap, so the same good can appear in bundles intended for different agents. Our construction ensures that each good appears in $O(\log^* n)$ bundles, improving the previous $O(\log\log n)$ bound~\cite{SS25}. We then apply Feige's rounding theorem~\cite{Feige25} to obtain disjoint bundles.
The smaller overlap bound reduces the loss from rounding and yields our improved MMS guarantee.
We explain the construction in the technical overview.

\paragraph{Constant approximations for $q$-partitioning valuations.}
Bangachev and Weinberg~\cite{BW25} recently introduced a hierarchy of $q$-partitioning valuations to study the classes between subadditive and XOS valuations. We use this hierarchy to ask what additional structure beyond subadditivity suffices for a constant MMS guarantee. A valuation is \emph{$q$-partitioning} if, for every set of goods and every partition of that set into at most $q$ groups, the groups admit nonnegative prices whose sum equals the set's value and whose sum on any collection of groups is at most the value of their union. These are supporting prices for the partition. The prices may depend on the partition, and the parameter $q$ bounds the number of groups for which they are required.
We denote this class by $\Q_q$. The class $\Q_2$ is precisely the subadditive class, whereas $\Q_m$ is the XOS class. The inclusions are strict before the hierarchy reaches XOS~\cite{BW25}. In this paper, we give the first study of MMS in the intermediate classes. Our second result gives a constant MMS approximation whenever $c=q/n$ is bounded below by a positive constant.

\begin{figure}
\centering
\begin{tikzpicture}
\begin{axis}[width=0.78\textwidth,height=5.2cm,xmin=0,xmax=2,ymin=0,ymax=.27,clip=false,
 xlabel={$c=q/n$},ylabel={MMS guarantee},
 xtick={0,1,1.2,2},xticklabels={$0$,$1$,$6/5$,$2$},
 ytick={0,.166666667,.25},yticklabels={$0$,$1/6$,$1/4$},
 grid=major,grid style={gray!15},axis line style={gray!60},tick label style={font=\normalsize}]
\addplot[very thick,black,domain=0:1,samples=100]{x/(11-5*x)};
\addplot[very thick,black,domain=1:1.2]{1/6};
\addplot[very thick,black,domain=1.2:2,samples=80]{2*(x-1)/(7*x-6)};
\addplot[only marks,mark=*,mark size=2pt,black] coordinates {(1,.166666667) (1.2,.166666667) (2,.25)};
\end{axis}
\end{tikzpicture}
\caption{The MMS guarantee in Theorem~\ref{thm:p-main}.}\label{fig:rho}
\end{figure}
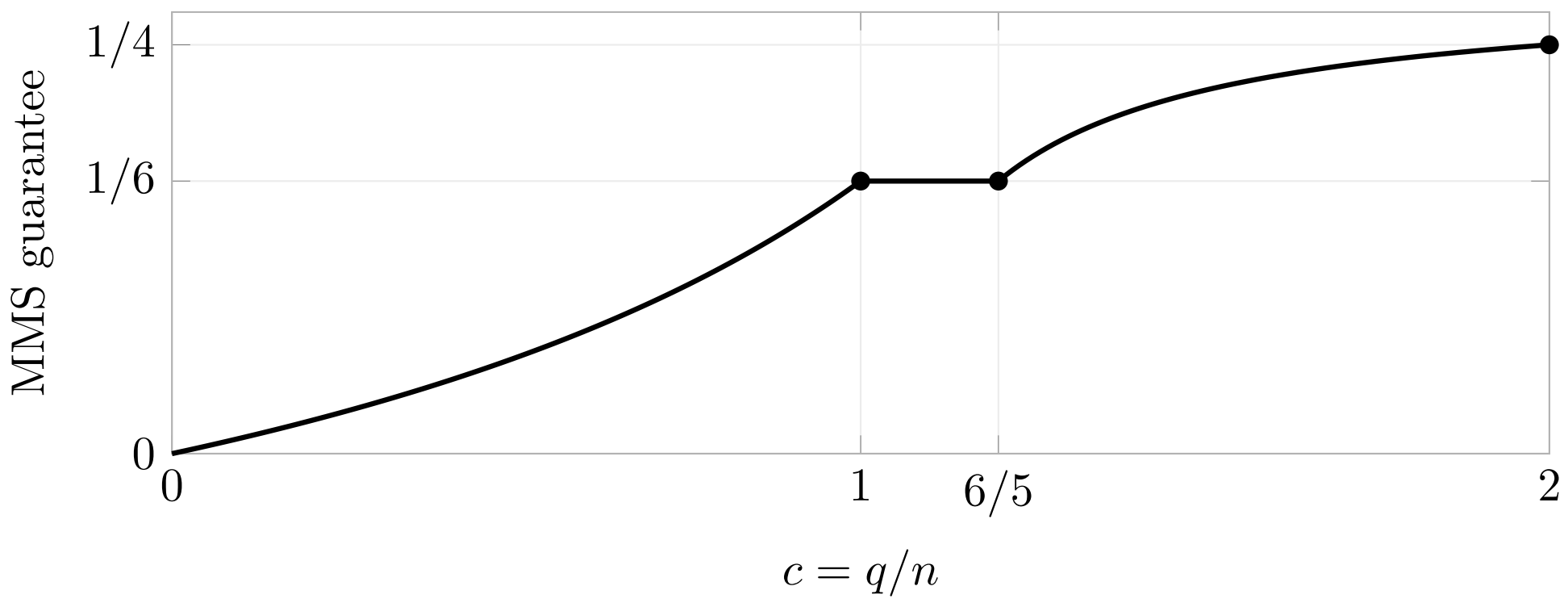

\begin{theorem}[Partitioning valuations]\label{thm:p-main}
For every integer $q\ge2$, every $n$-agent instance with valuations in $\Q_q$ admits a $\rho(q/n)$-MMS allocation, where
\begin{equation}\label{eq:rho}
\rho(c)=
\begin{cases}
\dfrac{c}{11-5c}, & 0<c\le1,\\[0.5em]
\dfrac16, & 1\le c\le\dfrac65,\\[0.5em]
\dfrac{2(c-1)}{7c-6}, & \dfrac65\le c\le2,\\[0.5em]
\dfrac14, & c\ge2.
\end{cases}
\end{equation}
\end{theorem}

Figure~\ref{fig:rho} shows how the guarantee $\rho(c)$ varies with $c=q/n$.
In particular, we obtain a $1/6$-MMS allocation when $q\ge n$ and a $1/4$-MMS allocation when $q\ge2n$.
Our theorem applies when $q=\Theta(n)$; when also $q<m$, $\Q_q$ strictly contains XOS, because its valuations need not admit supporting prices for every singleton partition.
It is worth noting that, for XOS valuations, this already gives a $1/4$ guarantee, improving the previous universal guarantee of $4/17$~\cite{FG25}.
To obtain the first two branches of~\eqref{eq:rho}, we group agents and round their allocations; for the last two, we transfer bundles directly to agents whose bundles do not yet meet the desired MMS guarantee.

\paragraph{A stronger guarantee for XOS valuations.}
We finally focus on the XOS endpoint of the hierarchy, and further improve the approximation factor.
\begin{theorem}[XOS valuations]\label{thm:xos-main}
Every instance with XOS valuations admits a $5/16$-MMS allocation.
\end{theorem}
The factor $5/16=0.3125$ improves the previous universal guarantee of $4/17$~\cite{FG25}. It also exceeds the recent guarantees of Feige and Grinberg~\cite{FG26}, which are above $11/40=0.275$ for sufficiently large $n$ and approach $0.2767738$ as $n$ grows. Our guarantee applies to all numbers of agents. The proof, given in Section~\ref{sec:xos-mms}, uses a reservation theorem that chooses selected goods assigned integrally together with auxiliary additive valuations on the remaining goods. This theorem accounts for the loss of one good in rounding; a cap-and-restore lemma makes the required fractional value small enough to obtain $5/16$.

Table~\ref{tab:results} summarizes our results and compares the known MMS guarantees in the literature.

\begin{table}
\centering
\begingroup
\setlength{\tabcolsep}{6pt}
\renewcommand{\arraystretch}{1.10}
\begin{tabular}{@{}>{\raggedright\arraybackslash}p{0.15\textwidth}>{\raggedright\arraybackslash}p{0.40\textwidth}>{\raggedright\arraybackslash}p{\dimexpr0.45\textwidth-4\tabcolsep\relax}@{}}
\toprule
\textbf{Valuations} & \textbf{Earlier guarantees} & \textbf{This paper}\\
\midrule
Subadditive & $\Omega(1/\log\log n)$~\cite{SS25,Feige25} & $\Omega(1/\log^* n)$ (Theorem~\ref{thm:sa-main})\\
\midrule[0.3pt]
$q$-partitioning & $\Omega(1/\log\log n)$~\cite{SS25,Feige25} &
constant when $q=\Theta(n)$ (Theorem~\ref{thm:p-main})\newline
$1/6$ when $q\ge n$\newline
$1/4$ when $q\ge2n$\\
\midrule[0.3pt]
XOS &
$4/17\approx 0.2353$ for all $n$~\cite{FG25}\newline
$11/40$ for sufficiently large $n$~\cite{FG26} &
$5/16=0.3125$ for all $n$ (Theorem~\ref{thm:xos-main})\\
\bottomrule
\end{tabular}
\endgroup
\caption{Comparison of MMS guarantees. The earlier bound for $q$-partitioning valuations follows from the general subadditive guarantee.}\label{tab:results}
\end{table}

\subsection[Technical overview]{Technical overview}
For each MMS guarantee, we first allocate any singleton that meets an agent's target and remove both the agent and the good, so that, after scaling, we may assume every agent has MMS one and values every singleton below the final target $\alpha$. We fix an MMS partition for each agent.

\paragraph{Subadditive valuations: bidding amplification.}
To prove \cref{thm:sa-main}, we construct bundles worth at least half the agents' MMS in successive \emph{layers}, where the allocated bundles within each layer are pairwise disjoint.
Once an agent receives a bundle worth at least half her MMS, we record that bundle and exclude her from subsequent layers, while keeping all goods available.
Every unserved agent therefore still has her $n$ disjoint MMS bundles.
Since each good appears in at most one recorded bundle per layer, the number of layers also bounds the number of bundles containing any particular good.
Thus, fewer layers reduce the overlap that must be resolved by rounding.

Our new technical contribution is a bidding amplification theorem.
We first consider an auxiliary bidding game in which a player submits nonnegative bids on goods, subject to a fixed budget, and wins each good on which her bid strictly exceeds the competing bid.
{Adapting the finite minimax argument of Neogi, Pashkovich, and Swamy~\cite{NPS24,NPS25},} we construct a distribution over bid vectors that guarantees probability at least \(1/2\) of winning goods worth at least half the value of a given bundle,
against every fixed competing bid vector whose total is at most half the player's budget. Every sampled bid vector satisfies the budget constraint.

We then use this strategy in multiple independent trials.
Suppose $u$ agents remain unserved.
Each unserved agent divides a unit budget among $\ell$ trials, independently choosing one of her $n$ MMS bundles uniformly at random and sampling a bid vector for the chosen bundle in each trial.
We sum each agent's trial bids coordinatewise to obtain her final bid vector, and assign every good to its unique highest bidder, if one exists.
Since all bids are nonnegative, every good strictly won by an individual trial against the opponents' final bids is also won by the agent's final bid.
Thus, one successful trial suffices for the agent to receive value at least half her MMS, and the probability that all trials fail decreases exponentially with the number of trials.
The following theorem formalizes this guarantee, and we present the proof in Section~\ref{sec:subadditive}.

\begin{theorem}[Amplification from disjoint valuable bundles]\label{thm:amplification}
Let $u\ge2$ and $n\ge1$ be integers. Suppose each of $u$ agents has a monotone, subadditive valuation that assigns zero to the empty set and $n$ pairwise disjoint sets of value at least one. Here $n$ counts the valuable sets available to each agent and need not equal $u$. For every integer $\ell\ge1$ with $2\ell(u-1)\le n$, there is a finitely supported distribution over feasible allocations such that, for each agent $i$,
\[
 \Prb[v_i(A_i)<1/2]\le
 \left(\frac12+\frac{\ell(u-1)}{n}\right)^\ell.
\]
If $n\ge4\ell(u-1)$, a feasible partial allocation leaves at most $\lfloor u(3/4)^\ell\rfloor$ agents below value $1/2$.
\end{theorem}

Let $u'$ denote the number of agents left unserved after one layer. The earlier guarantee~\cite{SS25} gives $u'\le u^2/n$. Once $n/u$ is sufficiently large and $u\ge2$, choosing $\ell$ proportional to $n/u$ in \cref{thm:amplification} instead gives
\[
 u'\le u\,2^{-n/(12u)}.
\]
Thus the guaranteed growth of $n/u$ improves from quadratic to exponential. Using the earlier guarantee for a few initial layers and amplification thereafter, we serve every agent in at most $1+\log_2^*n$ layers (Section~\ref{sec:amplified-layers}).
Finally, Feige's rounding theorem~\cite{Feige25}, together with the singleton bound, gives the $\Omega(1/\log^*n)$ MMS guarantee.

\paragraph{$q$-partitioning valuations: bounding losses from transfers.}
To prove \cref{thm:p-main}, we compare an agent's gain from receiving a bundle with the losses caused to the other agents, whom we call \emph{donors}. For an agent below the target, we consider transferring either a whole MMS bundle or one of two complementary pieces into which we split it. The transfer gives her its goods and removes them from the donors' bundles. We compare a disjoint family of candidates, evaluating each transfer separately from the same allocation.

Intersecting these candidates with a donor's bundle gives disjoint subsets. Supporting prices then bound their summed deletion loss: for a nonnegative integer $L$, a valuation $f\in\Q_q$, a bundle $S$, and pairwise disjoint subsets $B_1,\ldots,B_L\subseteq S$,
\begin{equation*}
 \sum_{\ell=1}^L\bigl(f(S)-f(S\setminus B_\ell)\bigr)
 \le\max\left\{1,\frac{L}{q}\right\}\,f(S).
\end{equation*}
Capping preserves $\Q_q$, so the same bound controls each donor's contribution to \emph{capped social welfare}, the sum of agents' bundle values capped at one. This potential is used in earlier XOS arguments~\cite{GHS+22,AMSS23,FG26}. With all agents and whole MMS bundles, however, the comparison guarantees an improving transfer only while the recipient's capped value is below $1/n$, even when $q\ge n$.

Our first construction divides the agents into groups and uses capped social welfare within each group. An agent still has all $n$ MMS bundles available, but fewer donors can lose value. More groups improve the guarantee within each group while increasing the rounding loss, since the group allocations use the same goods. Balancing these effects gives the first two branches of~\eqref{eq:rho}, including $1/6$ when $q\ge n$.

For $q>n$, our second construction keeps all agents together and changes the potential. Each agent contributes a bounded function of her bundle value that is zero below the target $\alpha$. We choose this function so that splitting an MMS bundle increases the guaranteed sum of its pieces' contributions, following the use of pieces in~\cite{AMSS23}. The difficulty is to bound donor losses: the function is concave only above $\alpha$, so the capping argument does not apply. We prove sufficient conditions for converting an aggregate bound on value losses into a bound on potential losses (Lemma~\ref{lem:potential-deletion}). Supporting prices supply the required aggregate bound.

The comparison uses at most $q$ disjoint candidates, choosing either each whole MMS bundle or its two pieces. Each split adds one candidate, so larger $q$ permits more splits and a larger recipient gain. When $q\ge2n$, splitting every bundle gives $1/4$. Both constructions repeatedly improve a bounded potential until all participating agents meet their target. Taking the better guarantee gives \cref{thm:p-main}; we give the proofs in Sections~\ref{sec:grouping} and~\ref{sec:direct}.

\paragraph{XOS valuations: a reservation theorem.}
To prove \cref{thm:xos-main}, we again use disjoint MMS bundles to bound the cost imposed by other agents, now through additive lower bounds on individual goods. We fix such a lower bound of total value one for each MMS bundle. We seek a fractional allocation under auxiliary additive valuations and round it following Shmoys and Tardos~\cite{ST93}. Rounding may lose the largest auxiliary coefficient. We therefore reserve selected goods for integral assignment and choose them together with an auxiliary valuation on the other goods and a threshold on fractional auxiliary value. We call this triple a choice; its threshold accounts for the loss in rounding.

Our reservation theorem (Theorem~\ref{xos:thm:reservation}) coordinates these choices. Against any fixed choices of the other agents with disjoint reserved sets, an agent must have a choice whose reserved goods avoid those sets and whose fractional auxiliary value exceeds the sum of its threshold and the cost of reserving its goods. We bound this cost by the other agents' auxiliary coefficients on those goods. To coordinate the choices, we use one independent mean-one exponential variable per agent. Our potential sums, over unreserved goods, the expected maximum of the agents' auxiliary coefficients multiplied by their variables, and subtracts the thresholds. The expected increase from adding an agent on a good equals her auxiliary coefficient times the probability that her product exceeds the others. These probabilities give feasible fractional shares. At a maximum of the potential, comparison with omitting an agent forces her fractional auxiliary value above her threshold, so rounding and adjoining the reserved goods meets her target.

For choices without reserved goods, we prove the cap-and-restore lemma (Lemma~\ref{xos:lem:cap}). We cap the coefficients of an additive lower bound and rescale them to sum to one. The capped coefficients control the rounding loss, while the original coefficients prove the target after rounding. When the original total value minus its largest coefficient exceeds the target, the sufficient fractional auxiliary value depends only on the target and this difference.

For $\alpha=5/16$, we first allocate pairs that meet an agent's target. We assign nonnegative credits to the surviving MMS bundles to account for the removed goods, with total credit at least the number of remaining agents. We bound each agent's auxiliary coefficient sum plus a fixed charge per reserved good by one. Disjointness then ensures that some bundle's credit exceeds the sum of the other agents' auxiliary coefficients on its unreserved goods and the charges for its goods reserved by those agents. We prove that such a bundle supplies a suitable choice. In the remaining local case, a weighted combination of lower bounds on the potential increases for three choices cancels the other agents' coefficients on the two goods with largest coefficients in the original additive lower bound. At least one actual choice therefore works. This gives a $5/16$-MMS allocation; we give the proof in Section~\ref{sec:xos-mms}.

\subsection[Related work]{Related work}\label{sec:related}

\paragraph{MMS guarantees for additive, submodular, and XOS valuations.}
Budish~\cite{Budish11} introduced the MMS benchmark for the allocation of indivisible goods. For additive valuations, Kurokawa, Procaccia, and Wang~\cite{PW14} proved that a $2/3$-MMS allocation always exists; subsequent improvements~\cite{GHS+18,AG24,HKSS26,HZ25} have raised the guarantee to $7/9$~\cite{HZ25}. For submodular valuations, the initial $1/10$ guarantee of Barman and Krishnamurthy~\cite{BK17} was improved to $1/3$ by Ghodsi et al.~\cite{GHS+22} and then to $10/27$ by Ben Uziahu and Feige~\cite{BUF23}. Ghodsi et al.~\cite{GHS+22} also proved constant guarantees for XOS valuations, followed by improvements~\cite{SS22,AMSS23,FG25,FG26}.
Feige and Grinberg~\cite{FG25,FG26} prove their XOS guarantees for the stronger anyprice share (APS) benchmark and hence also for MMS.
Informally, APS is the largest value an agent can guarantee with budget $1/n$ under any nonnegative prices summing to one.
Related work also considers hereditary set systems~\cite{LV21,Hummel25}, matroid-rank valuations~\cite{BV21}, and XOS valuations with marginal values in $\{0,1\}$~\cite{KKM24,CLLZ26}.
We refer to Amanatidis et al.~\cite{AAB+23} for a survey of
fair division of indivisible goods.

\paragraph{General subadditive valuations.}
Ghodsi et al.~\cite{GHS+22} obtained an $\Omega(1/\log m)$ guarantee. Seddighin and Seddighin~\cite{SS22} replaced the dependence on the number of goods with a dependence on the number of agents, and Feige and Huang~\cite{FH25} improved the guarantee to $1/(14\log n)$. Seddighin and Seddighin~\cite{SS25} then constructed half-MMS bundles using $O(\log\log n)$ copies of each good. Combining this construction with Feige's rounding theorem~\cite{Feige25}, whose loss is linear in the number of copies, gives an $\Omega(1/\log\log n)$ guarantee. We use the same rounding step and reduce the number of copies.
Even two subadditive agents can rule out guarantees above $1/2$~\cite{GHS+22}. Christodoulou et al.~\cite{CCMS25} attain $1/2$ for at most four agents and for any number of agents with only two distinct valuations. Feige~\cite{Feige26BoBW} studies a related direction: randomized allocations that preserve an MMS guarantee in every outcome and give additional guarantees in expectation relative to the best average value of a partition into $n$ bundles.
For MMS guarantees for chores with nonadditive costs, we refer to
Li, Wang, and Zhou~\cite{LWZ23}.

\paragraph{Bidding and amplification.}
Neogi, Pashkovich, and Swamy~\cite{NPS24,NPS25} develop randomized bidding strategies with expected-value guarantees and budgets that hold in every outcome, using finite minimax arguments and random selection. We adapt this minimax framework to an indicator payoff for winning half a bundle's value and enforce strict wins by a grid construction. A bidding-based proof of the earlier partial-allocation result~\cite{SS25} appears in~\cite{MNPS+26}, using samples of opponents' bids and their coordinatewise maxima. Our amplification instead combines independent trials of an agent's own strategy with coordinatewise summation, so that one successful trial suffices and every final bid obeys its budget. Conditioning on the opponents' final bids makes the trial successes independent and gives the exponential failure bound in \cref{thm:amplification}.

\paragraph{Partitioning valuations and capped welfare.}
Bangachev and Weinberg~\cite{BW25} introduced the partitioning hierarchy and its supporting-price characterization. Earlier XOS arguments use capped social welfare~\cite{GHS+22,AMSS23,FG26}, complementary pieces~\cite{AMSS23}, and bounds on losses from disjoint deletions~\cite{GHS+22}. We extend the deletion bound to $q$-partitioning valuations and prove a scalar criterion that controls losses after applying a function that is zero below the target. This criterion permits the potential in our second construction to strengthen the gain from pieces while bounding each donor's summed loss.
Our stronger XOS result uses additive assignment rounding with a loss of at most one good, following Shmoys and Tardos~\cite{ST93}. The reservation theorem chooses the goods assigned integrally together with the auxiliary coefficients used in this rounding step. The cap-and-restore lemma and the combination of local reservation choices account for large coefficients and competing demands on the same goods. Unlike the hierarchy proof, this argument uses additive lower bounds on arbitrary subsets of an MMS bundle.

\section{Preliminaries}\label{sec:preliminaries}
\subsection{Model and solution concepts}
We consider a finite set $M$ of $m$ goods and a set $[n]=\{1,\ldots,n\}$ of agents. Each agent $i$ has a valuation function $v_i:2^M\to\R_{\ge0}$. Every valuation considered here is \emph{normalized}, i.e., $v_i(\varnothing)=0$, and \emph{monotone}, i.e., $v_i(S)\le v_i(T)$ for $S\subseteq T$. A valuation $v$ is \emph{subadditive} if, for all $S,T\subseteq M$,
\[
 v(S\cup T)\le v(S)+v(T).
\]
A bundle is a subset of goods, and a singleton is a bundle containing one good. An allocation is a tuple of pairwise disjoint bundles $(A_i)_{i\in[n]}$; this disjointness is what we mean by {\em feasibility} in later technical sections. An allocation is complete if the bundles cover $M$. A partial allocation may leave goods unassigned and agents with empty bundles.
By monotonicity, any partial allocation can be completed by assigning the remaining goods arbitrarily, without decreasing any agent’s value.

The MMS benchmark evaluates a partition from the perspective of the agent who receives its least valuable bundle. In order to apply allocation results to auxiliary instances, we allow the number of bundles and the available goods to vary in our notation. For a subset $T\subseteq M$ and a positive integer $k$, we let $\Pi(T,k)$ denote the set of all ordered partitions $(P_1,\ldots,P_k)$ of $T$ into $k$ possibly empty bundles. We define
\[
 \MMS_k(v;T)=\max_{(P_1,\ldots,P_k)\in\Pi(T,k)}
                  \min_{j\in[k]}v(P_j).
\]
The inner minimum is the guarantee of a particular partition; the outer maximum chooses the best such guarantee. We denote agent $i$'s MMS benchmark by $\mu_i=\MMS_n(v_i;M)$. An \emph{MMS partition} for agent $i$ is a partition in $\Pi(M,n)$ attaining $\mu_i$, and we call each bundle in such a partition an \emph{MMS bundle}. An $\alpha$-MMS allocation gives every agent $i$ a bundle of value at least $\alpha\mu_i$.
Without loss of generality, we consider only agents with $\mu_i>0$: agents with $\mu_i=0$ meet their target with the empty bundle, and removing them cannot lower the remaining agents' MMS values.

All logarithms are base two except $\ln$. A \emph{dyadic} positive integer is a power of two. For the iterated logarithm, we define
\begin{equation}\label{eq:towers}
\begin{aligned}
 E_0&=1,\\
 E_{r+1}&=2^{E_r},\\
 \log_2^* n&=\min\{r\ge0:n\le E_r\}.
\end{aligned}
\end{equation}
For a vector $b$ indexed by goods, let $b(S)=\sum_{g\in S}b_g$, and let $x_+=\max\{x,0\}$.

\subsection{The partitioning hierarchy}\label{sec:hierarchy}

An additive valuation assigns a nonnegative value $a_g$ to each good $g$ and values a bundle by summing the values of its goods.
An XOS valuation $v$ is the maximum of finitely many additive valuations. Equivalently, for every $S\subseteq M$, there are nonnegative {\em supporting prices} on its goods that sum to $v(S)$ and whose total on any subset $T\subseteq S$ is at most $v(T)$.
{Extending these prices by zero outside $S$ defines an additive valuation $w$ satisfying $w(S)=v(S)$ and $w(T)\le v(T)$ for every $T\subseteq M$.}

Bangachev and Weinberg~\cite{BW25} introduced the hierarchy of $q$-partitioning valuations to study classes strictly between XOS and subadditive valuations. The idea is to relax the XOS supporting-price requirement by grouping goods. For every partition of a bundle into at most $q$ groups, we require supporting prices on the groups: each price applies to a whole group, and the total price of any collection of groups is at most the value of their union. The groups may contain arbitrarily many goods. Increasing $q$ requires prices for more partitions, up to the XOS requirement at $q=m$. We use their supporting-price characterization~\cite{BW25} as our definition.

\begin{definition}[$q$-partitioning valuation]\label{def:partitioning}
Let $q\ge2$ be an integer. A normalized monotone valuation $v$ belongs to $\Q_q$ if, for every $S\subseteq M$ and every partition $(C_1,\ldots,C_t)\in\Pi(S,t)$ into $t\le q$ nonempty groups, there are nonnegative values $w_1,\ldots,w_t$ such that, for every $J\subseteq[t]$,
\begin{equation}\label{eq:group-prices}
\begin{aligned}
 \sum_{j=1}^t w_j&=v(S),\\
 \sum_{j\in J}w_j&\le v\!\left(\bigcup_{j\in J}C_j\right).
\end{aligned}
\end{equation}
The values $w_j$ are supporting prices for this partition.
\end{definition}

The equality in~\eqref{eq:group-prices} distributes the whole set's value among its $t$ groups. Each inequality ensures that the prices do not overstate the value of a collection of groups. In particular, the groups other than $C_j$ have total price $v(S)-w_j$. By applying the inequality to those groups, we obtain $v(S\setminus C_j)\ge v(S)-w_j$, or equivalently
\[
 v(S)-v(S\setminus C_j)\le w_j.
\]
Thus the price of a group bounds the value lost by deleting it. In our transfer arguments in Section \ref{sec:partitioning}, we will use supporting prices to sum these deletion bounds. The prices may change with the set and its partition; no common prices across different partitions are required. We use the at-most-$q$ convention also when $q>m$, in which case the class is XOS.

For $q=2$, monotonicity and subadditivity give the required prices for a partition $(C_1,C_2)\in\Pi(S,2)$: we may take $w_1=v(C_1)$ and $w_2=v(S)-v(C_1)$. At $q=m$, singleton partitions recover XOS supporting prices; conversely, summing these prices within each group gives supporting prices for every partition. Bangachev and Weinberg~\cite{BW25} proved that, for $m\ge2$, these classes form the hierarchy
\[
 \{\text{XOS valuations}\}=\Q_m\subseteq\cdots\subseteq\Q_2
 =\{\text{subadditive valuations}\}.
\]
{Bangachev and Weinberg~\cite{BW25} proved that the inclusions in this hierarchy are strict whenever $2\le q<m$.
In particular, \(\mathcal Q_q\) strictly contains XOS for every $q<m$.
Consequently, our constant MMS guarantees for $q=\Theta(n)$ also apply to valuations beyond XOS.}

\subsection{\texorpdfstring{Allocating singletons and making bundles disjoint}{Allocating singletons and making bundles disjoint}}

We use the standard singleton reduction~\cite{GHS+22,AMSS23} to assume that every singleton has value below the target. This bound controls the loss in rounding.

\begin{lemma}\label{lem:singleton-reduction}
For $n\ge2$, allocating a good $g$ to an agent $i$ and removing both ensures that, for every $j\ne i$,
\[
 \MMS_{n-1}(v_j;M\setminus\{g\})\ge\mu_j.
\]
In particular, if $v_i(\{g\})\ge\alpha\mu_i$, this allocation meets agent $i$'s $\alpha$-MMS target without lowering any remaining agent's MMS.
\end{lemma}
\begin{proof}
For every $j\ne i$, take an MMS partition of $M$ for agent $j$. Discard the bundle containing $g$ and distribute its goods other than $g$ among the remaining $n-1$ bundles. These bundles partition $M\setminus\{g\}$ and each has value at least $\mu_j$ by monotonicity.
\end{proof}

Whenever a singleton meets an agent's MMS approximation target, we may allocate it to that agent and remove both.
By Lemma~\ref{lem:singleton-reduction}, the remaining agents' MMS benchmarks do not decrease.
Moreover, the approximation factors obtained in this paper do not decrease when the number of agents decreases, with $q$ fixed for the partitioning results.
Consequently, an allocation satisfying the guarantee for the reduced instance also satisfies the original targets of the remaining agents.
We may repeatedly apply this reduction, with the case of a single remaining agent handled by assigning all remaining goods to her.

It thus suffices to prove each guarantee when no singleton meets its target. We then divide each valuation by its MMS and reuse $M$, $n$, and $v_i$ for this working instance. Restriction to the remaining goods and positive scaling preserve both subadditivity and membership in $\Q_q$. For an MMS target $\alpha$, our allocation constructions hence assume that, for every agent $i\in[n]$ and good $g\in M$,
\[
 \MMS_n(v_i;M)=1 \text{ and } v_i(\{g\})<\alpha.
\]
We fix an MMS partition $(W_{i,1},\ldots,W_{i,n})\in\Pi(M,n)$ for each agent. Every $W_{i,j}$ has value at least one. These assumptions apply to the final MMS allocation constructions; our intermediate results for disjoint valuable bundles apply as stated.

\paragraph{Overlapping allocations and rounding.}
In our constructions, we sometimes obtain several feasible partial allocations on the same set of goods and select one bundle for each agent from these allocations.
These selected bundles may overlap.
For a positive integer $d$, a \emph{$d$-multi-allocation} is a tuple $(B_i)$ in which each good belongs to at most $d$ bundles. Its \emph{multiplicity} is the largest number of bundles containing a single good. A \emph{layer} is a feasible partial allocation on the same goods.
If each agent selects her bundle from one of at most $d$ layers, the resulting tuple is a $d$-multi-allocation, since a good occurs at most once per layer. We keep the goods available when constructing another layer and use the final rounding step to resolve their competing uses.

In our subadditive construction, we serve agents in successive layers. Let $u\le n$ be the number still unserved, while $n$ and each unserved agent's MMS partition stay fixed.
For $q$-partitioning valuations, we divide agents into groups and construct a feasible allocation for each group on the same set of goods.
For both constructions, we use the following theorem to turn the overlapping bundles into a feasible allocation. Constructions that maintain feasibility throughout do not require this rounding step.

\begin{theorem}[Feige's subadditive rounding theorem~{\cite{Feige25}}]\label{thm:rounding}
Let $D\ge2$ be a power of two. For any $D$-multi-allocation $(B_i)$ under monotone subadditive valuations, let $\delta_i=\max_{g\in B_i}v_i(\{g\})$, with $\delta_i=0$ when $B_i=\varnothing$. There are pairwise disjoint sets $A_i\subseteq B_i$ such that, for every agent $i$,
\begin{equation}\label{eq:rounding}
 v_i(A_i)\ge\frac{v_i(B_i)}{D}-\frac{D-1}{D}\delta_i.
\end{equation}
\end{theorem}

The first term in~\eqref{eq:rounding} divides the intermediate bundle's value by the multiplicity bound $D$. The second is the loss from indivisibility, controlled by its largest singleton value. This singleton value is less than $\alpha$. If every intermediate bundle has value at least $\eta$, it therefore suffices to ensure
\[
 \frac{\eta-(D-1)\alpha}{D}\ge\alpha,
 \qquad\text{equivalently}\qquad
 \alpha\le\frac{\eta}{2D-1}.
\]
We use this relation to choose the final target $\alpha$ from the intermediate guarantee.
In the subadditive construction, we use $\eta=1/2$, whereas in the grouping construction, the intermediate guarantee $\eta$ depends on the number of groups.
In both constructions, we choose a power of two $D\ge2$ at least as large as the multiplicity of the resulting multi-allocation.

\section{Subadditive valuations: exponential amplification}\label{sec:subadditive}
We prove \cref{thm:sa-main} in three stages. We first construct a randomized bidding strategy for a single set of goods using a finite minimax argument.
We then apply this strategy in multiple independent trials to amplify the probability that an agent receives goods of value at least $1/2$.
We next apply the amplification theorem to the working instance, using each agent's fixed MMS partition to construct successive layers of feasible partial allocations.
We show that at most $1+\log_2^*n$ layers suffice to give every agent value at least $1/2$.
Since each layer is feasible, the resulting multi-allocation has multiplicity at most $1+\log_2^*n$.
Finally, Feige's rounding theorem, together with the singleton bound, gives the desired MMS guarantee.

\subsection{A strategy for winning half of one bundle}
We first consider a bidding problem on a nonempty finite set $P$ of goods. A bid vector $b$ assigns a nonnegative number $b_g$ to each good $g\in P$.
Against a competing bid vector $c$, the bid vector $b$ strictly wins a good $g\in P$ if $b_g>c_g$. Here, bids are used only to determine which goods are won and do not represent monetary payments.

Given a valuation $v$ on $P$, we seek a distribution over bid vectors that does not depend on the competing bids and satisfies a prescribed budget in every realization.
Against every fixed competing bid vector whose total is at most half this budget, the distribution should guarantee probability at least $1/2$ of winning goods worth at least $v(P)/2$.
This probability guarantee, rather than a bound on expected value, is needed for the independent trials in the subsequent construction. The following lemma proves these properties using a finite minimax argument.

\begin{lemma}\label{lem:minimax}
Let $P\subseteq M$ be a nonempty finite set, let $v:2^P\to\R_{\ge0}$ be a normalized, monotone, subadditive valuation, and let $B>0$.
There is a finitely supported distribution $\sigma$ over nonnegative bid vectors on $P$ such that $b(P)\le B$ for every $b\in \supp(\sigma)$, and for every fixed nonnegative bid vector $c$ on $P$ with $c(P)\le B/2$,
\begin{equation*}
 \Prb_{b\sim \sigma}\!\left[v(\{g\in P:b_g>c_g\})\ge\tfrac12v(P)\right]\ge\tfrac12,
\end{equation*}
where $\supp(\sigma)$ is the support of $\sigma$.
\end{lemma}

\begin{proof}
We first consider a finite bidding game in which bids are integer multiples of a fixed amount and ties count as wins.
By the minimax theorem, we obtain a distribution over these bids.
We then increase each sampled bid by the fixed amount to guarantee strict wins against arbitrary nonnegative real competing bids with the required probability.

\paragraph{A finite game on a grid.}
Let $s=|P|$ and $h=B/(2s)$.
The grid spacing $h$ is chosen so that increasing every coordinate by $h$ increases the total bid by exactly $B/2$.
Let
\[
 K=\left\{hz:z\in\mathbb Z_{\ge0}^{P},\ \sum_{g\in P}z_g\le s\right\}.
\]
Thus $K$ consists of all nonnegative \emph{grid} bids whose coordinates are integer multiples of $h$ and whose total is at most $B/2$. This set is finite. For any two bid vectors $d,c\in K$, define
\[
 H(d,c)=\one\!\left[v(\{g\in P:d_g\ge c_g\})\ge\tfrac12v(P)\right].
\]
Here $\one$ is the indicator. We regard $H$ as the payoff matrix of a finite game, with the row player maximizing $H(d,c)$ and the column player minimizing it.

For any $d,c\in K$ and any $g\in P$, either $d_g\ge c_g$ or $c_g\ge d_g$, and both hold in a tie. Hence the set weakly won by $d$ against $c$ and the set weakly won by $c$ against $d$ together cover $P$. By the subadditivity of $v$, we have
\[
 v(P)\le v(\{g\in P:d_g\ge c_g\})+v(\{g\in P:c_g\ge d_g\}).
\]
The two terms cannot both be strictly below $v(P)/2$. Therefore, for any $d,c\in K$,
\begin{equation}\label{eq:symmetric-payoff}
 H(d,c)+H(c,d)\ge1.
\end{equation}

\paragraph{One distribution against every grid bid.}
Let $\gamma$ be an arbitrary probability distribution on $K$, and sample bid vectors $d$ and $c$ independently from $\gamma$.
Swapping their names does not change their joint distribution, so
\[
\E_{d,c\sim\gamma}H(d,c) = \E_{d,c\sim\gamma}H(c,d).
\]
Averaging~\eqref{eq:symmetric-payoff} then gives $\E_{d,c\sim\gamma}H(d,c)\ge1/2$.
In particular, there exists some fixed row strategy $d^*\in K$ that has expected payoff at least $1/2$ against a column strategy sampled from $\gamma$.

Let $\Delta(K)$ denote the simplex of probability distributions on $K$. By von Neumann's minimax theorem,
\[
\max_{\pi\in \Delta(K)}\min_{c\in K}\E_{d\sim\pi}H(d,c)
 =\min_{\gamma\in \Delta(K)}\max_{d\in K}\E_{c\sim\gamma}H(d,c).
\]
The right-hand side is at least $1/2$ by the preceding paragraph.
Fix a maximizing distribution $\pi^*$ on the left-hand side. It follows that, for each bid vector $c\in K$,
\[
 \E_{d\sim\pi^*}H(d,c)\ge\tfrac12.
\]

\paragraph{Strict wins against arbitrary real bids.}
Finally, based on $\pi^*$, we construct the probability distribution $\sigma$ over bid vectors on $P$ that satisfies the conditions in the lemma statement.
Recall that $s=|P|$ and \(h=B/(2s)\). To construct \(\sigma\), sample a bid vector \(d\sim\pi^*\) and define \(b_g=d_g+h\) for every \(g\in P\). Let \(\sigma\) denote the probability distribution of the resulting vector \(b\). Since \(\pi^*\) is a distribution on the finite set \(K\), the distribution \(\sigma\) has finite support. Moreover, every \(d\in K\) satisfies \(d(P)\le B/2\), and adding \(h\) to each of the \(s\) coordinates increases the total bid by \(sh=B/2\). Consequently, every \(b\in\supp(\sigma)\) satisfies \(b(P)\le B\).

It remains to verify the strict winning guarantee. Fix an arbitrary nonnegative real bid vector \(c\) on \(P\) with \(c(P)\le B/2\). Notice that $c\notin K$ is possible.
For each \(g\in P\), define \(c'_g=h\lfloor c_g/h\rfloor\).
Since rounding down cannot increase the total, we have $c'\in K$.
Moreover, by the definition of $c'$, we have \(c_g<c'_g+h\) for every $g\in P$. Therefore, for every \(g\in P\), if \(d_g\ge c'_g\), then
\[
b_g=d_g+h\ge c'_g+h>c_g.
\]
It follows that \(\{g\in P:d_g\ge c'_g\}\subseteq\{g\in P:b_g>c_g\}\). By the monotonicity of \(v\), the value of the set strictly won by \(b\) against \(c\) is at least the value of the set weakly won by \(d\) against \(c'\).
Since \(c'\in K\), the guarantee for \(\pi^*\) implies
\[
\begin{aligned}
&\Prb_{b\sim\sigma}\left[
v(\{g\in P:b_g>c_g\})\ge\frac{v(P)}{2}
\right]\\
&\quad\ge
\Prb_{d\sim\pi^*}\left[
v(\{g\in P:d_g\ge c'_g\})\ge\frac{v(P)}{2}
\right]\\
&\quad=
\mathbb E_{d\sim\pi^*}H(d,c')
\ge\frac12.
\end{aligned}
\]
Hence, the same distribution \(\sigma\) satisfies the required probability bound for every nonnegative real vector \(c\) with \(c(P)\le B/2\). This completes the proof.
\end{proof}

\paragraph{Constructing the bidding distribution.}
Given $P,v,B$, construct $K$ and $H$ as in the proof and solve the finite system of linear inequalities
\[
 \pi\in\Delta(K),\qquad
 \sum_{d\in K}\pi(d)H(d,c)\ge\frac12\quad \text{for every }c\in K.
\]
The minimax argument guarantees that this system has a solution.
Let $s=|P|$ and $h=B/(2s)$.
Choose such a distribution $\pi$, and let $\sigma$ be the distribution of the bid vector $b$ obtained by sampling \(d\sim\pi\) and setting \(b_g=d_g+h\) for every \(g\in P\).
We denote the procedure that returns $\sigma$ by $\textsc{BidDistribution}(P,v,B)$.
It depends only on $P,v,B$, and not on the competing bid vector.
Since $|K|=\binom{2s}{s}$, this finite construction may require exponential time in $s$.

\subsection{Independent trials with one total budget}
We prove \cref{thm:amplification} by applying the bidding strategy of Lemma~\ref{lem:minimax} in multiple independent trials under a single unit budget.
In each trial, an agent uniformly at random chooses one of her $n$ pairwise disjoint sets of value at least one and samples a bid vector for that set.
To combine these trials into one allocation, we sum each agent's trial bids coordinatewise to obtain her final bid vector.
Since all bids are nonnegative, every good strictly won by an individual trial against the opponents' final bids is also won by the agent's final bid.
Moreover, since these $n$ sets are pairwise disjoint, the opponents cannot place large competing bids on too many of them.
This allows us to bound the probability that all trials fail.

\paragraph{The repeated-bidding procedure.}
Its input consists of \(u\ge2\) agents with normalized, monotone, subadditive valuations \(v_1,\ldots,v_u\), \({n}\ge1\) pairwise disjoint sets \(W_{i,1},\ldots,W_{i,{n}}\) for each agent \(i\in[u]\), with \(v_i(W_{i,r})\ge1\) for every \(r\in[{n}]\), an integer \(\ell\ge1\) satisfying \(2\ell(u-1)\le {n}\), and, for each \(i\in[u]\) and \(r\in[n]\), the probability distribution
\[
\sigma_{i,r}=\textsc{BidDistribution}(W_{i,r},v_i,1/\ell).
\]
Each agent performs \(\ell\) independent trials. In each trial, she selects an index \(r\in[{n}]\) uniformly at random and samples a bid vector from the corresponding distribution \(\sigma_{i,r}\), assigning zero bids to goods outside \(W_{i,r}\).
The random choices made in different trials, including those made by different agents, are independent.
For each agent $i\in[u]$ and trial $t\in[\ell]$, let $b_i^{(t)}$ denote her sampled bid vector, and let $b_i=\sum_{t=1}^\ell b_i^{(t)}$ be her final bid vector. By Lemma~\ref{lem:minimax}, each trial has total bid at most $1/\ell$, so $b_i(M)\le1$ in every realization. We assign each good to its unique highest bidder, leaving it unassigned if the highest bid is tied. Let $(A_i)_{i\in[u]}$ denote the resulting allocation, which is feasible in every realization. We denote this randomized procedure by $\textsc{RepeatedBidding}$.

When the stronger condition \({n}\ge4\ell(u-1)\) holds, we enumerate the finitely many possible realizations of the randomized procedure and choose one in which at most \(\lfloor u(3/4)^\ell\rfloor\) agents receive value below \(1/2\). We then replace the bundles of these agents by empty sets, leaving their goods unassigned.
In this case, $\textsc{RepeatedBidding}$ returns a feasible partial allocation in which at most \(\lfloor u(3/4)^\ell\rfloor\) agents remain unserved. The proof of Theorem~\ref{thm:amplification} below proves that such a realization exists, so this finite search terminates.
We will use this stronger guarantee to construct the amplified layers in Section~\ref{sec:amplified-layers}.

\begin{proof}[Proof of \cref{thm:amplification}]


We first analyze the randomized allocation $(A_i)_{i\in[u]}$ before the outcome-selection step.

\paragraph{The success probability of one trial.}
Fix an agent $i$ and condition on the final bid vector of all agents $j\neq i$.
Under this conditioning, for each agent $j\ne i$ and each good \(g\in M\), the bid \(b_{j,g}\) is fixed.
For each good \(g\in M\), let \(c_g=\max_{j\ne i}b_{j,g}\) denote the largest competing bid. By the allocation rule, agent \(i\) wins good \(g\) if and only if her final bid exceeds \(c_g\).
Since all bids are nonnegative, we have \(c_g\le\sum_{j\ne i}b_{j,g}\) for every $g$. Summing over all goods and recalling that \(b_j(M)\le1\) for every agent \(j\), we obtain
\[
c(M)=\sum_{g\in M}c_g
\le\sum_{j\ne i}\sum_{g\in M}b_{j,g}
=\sum_{j\ne i}b_j(M)
\le u-1.
\]

Recall that \(W_{i,1},\ldots,W_{i,{n}}\) are \({n}\) pairwise disjoint sets of value at least one for agent \(i\). For each \(k\in[{n}]\), the quantity \(c(W_{i,k})\) represents the total competing bid on the goods in \(W_{i,k}\). Since these sets are pairwise disjoint and all competing bids are nonnegative, each coordinate \(c_g\) is counted in at most one of their totals. Consequently,
\[
\sum_{k=1}^{{n}}c(W_{i,k})\le c(M)\le u-1.
\]

We claim that at least \({n}-2\ell(u-1)\) of these sets satisfy \(c(W_{i,k})\le 1/(2\ell)\). Indeed, if more than \(2\ell(u-1)\) sets had competing total strictly greater than \(1/(2\ell)\), the sum of their competing totals would exceed \(u-1\), contradicting the preceding inequality.
Thus, at least a fraction \(1-2\ell(u-1)/{n}\) of the \({n}\) sets have competing total at most \(1/(2\ell)\). By the assumption \(2\ell(u-1)\le {n}\), this fraction is nonnegative.

Now consider an arbitrary trial \(t\in[\ell]\) of agent \(i\). Recall that the trial selects one of \(W_{i,1},\ldots,W_{i,{n}}\) uniformly at random and then samples a bid vector from the distribution provided by Lemma~\ref{lem:minimax}, for the selected set with budget \(1/\ell\).
By the preceding argument, the probability that the selected set \(W\) satisfies \(c(W)\le 1/(2\ell)\) is at least \(1-2\ell(u-1)/{n}\). Conditional on selecting such a set \(W\), the competing bid vector $c$ restricted to \(W\) is nonnegative and has total at most \(1/(2\ell)\).
Therefore, Lemma~\ref{lem:minimax} implies that, with probability at least \(1/2\), the randomized bid vector \(b_i^{(t)}\) sampled in this trial satisfies
\[
v_i\bigl(\{g\in W:b_{i,g}^{(t)}>c_g\}\bigr)
\ge \frac{v_i(W)}{2}
\ge \frac12,
\]
where the last inequality follows from \(v_i(W)\ge1\).

We call trial \(t\) \emph{successful} if the set of goods on which its bid vector strictly exceeds the competing bid vector \(c\) has value at least \(1/2\).
Since the bid vector in each trial is zero outside its selected set, a successful trial must win goods worth at least \(1/2\) within that set.
Combining the probability of selecting a set with competing total at most \(1/(2\ell)\) and the conditional success probability guaranteed by Lemma~\ref{lem:minimax}, we conclude that the probability of success in each trial, conditional on the opponents' final bid vectors, is at least
\[
\theta
=
\frac12\left(1-\frac{2\ell(u-1)}{{n}}\right).
\]
Notice that this lower bound holds for every possible realization of the competing final bid vectors, since the only property of these vectors used in the argument is that their total competing bid is at most \(u-1\).

\paragraph{From trial success to the allocation guarantee.}
The opponents' final bids depend only on their own randomness, so conditioning on $c$ leaves agent $i$'s trials independent with their original distributions. Each trial's success depends only on that trial and the fixed vector $c$. Let $S_t$ denote the event that trial $t$ succeeds. These events are therefore conditionally independent, with $\Prb(S_t\mid c)\ge\theta$.

If trial $t$ succeeds, its strict-winning set has value at least $1/2$. Since $b_i=\sum_{t=1}^{\ell}b_i^{(t)}\ge b_i^{(t)}$ coordinatewise, every good in this set is also strictly won by the final bid and belongs to $A_i$. Monotonicity then gives $v_i(A_i)\ge1/2$. Consequently, failure in the final allocation requires every trial to fail, and
\[
 \Prb\!\left(v_i(A_i)<\frac12\,\middle|\,c\right)
 \le\prod_{t=1}^{\ell}\bigl(1-\Prb(S_t\mid c)\bigr)
 \le(1-\theta)^\ell.
\]
Since this bound holds for every realization of $c$, averaging over $c$ gives
\[
 \Prb\!\left(v_i(A_i)<\frac12\right)
 \le(1-\theta)^\ell
 =\left(\frac12+\frac{\ell(u-1)}n\right)^\ell,
\]
as claimed.

\paragraph{A partial allocation with few unserved agents.}
Suppose further that \({n}\ge 4\ell(u-1)\). Then \(\frac12+\frac{\ell(u-1)}{{n}}\le\frac34\), and hence, by the preceding probability bound, we have \(\Prb[v_i(A_i)<1/2]\le(3/4)^\ell\) for every agent \(i\in[u]\).
Let \(F\) denote the number of agents whose allocated bundles have value strictly less than \(1/2\). By the linearity of expectation, we have
\[
\mathbb E[F]
=\sum_{i=1}^{u}\Prb\left(v_i(A_i)<\frac12\right)
\le u\left(\frac34\right)^\ell.
\]
Since $F$ takes integer values, some outcome has at most $\lfloor u(3/4)^\ell\rfloor$ agents below value $1/2$.

Fix such a realization and consider its resulting feasible allocation \((A_i)_{i\in[u]}\). We keep the bundle \(A_i\) of every agent \(i\) satisfying \(v_i(A_i)\ge1/2\), and replace the bundles of all other agents by the empty set, leaving some goods unassigned.
Since the original allocation is feasible, the resulting partial allocation remains feasible.
Moreover, every agent receiving a nonempty bundle has value at least \(1/2\), and at most \(\lfloor u(3/4)^\ell\rfloor\) agents remain unserved. This proves the theorem.
\end{proof}

\subsection{From amplification to an iterated logarithm}\label{sec:amplified-layers}
We now use the partial allocations constructed in the preceding section to obtain a multi-allocation in which every agent receives value at least \(1/2\).
We construct a sequence of feasible partial allocations, called layers. In each layer, we consider only agents who have not received value at least \(1/2\) in any previous layer.
We record the bundles of agents who meet this threshold and exclude these agents from subsequent layers. Goods assigned in one layer remain available in later layers, so every unserved agent continues to have her \({n}\) pairwise disjoint sets of value at least one.
Since each layer is feasible, every good belongs to at most one recorded bundle from that layer.
Therefore, the multiplicity of the resulting multi-allocation is at most the number of layers.

Our goal is to serve all agents using at most \(1+\log_2^*{n}\) layers. Theorem~\ref{thm:amplification} becomes more effective as the number of unserved agents decreases, because each agent can perform more independent bidding trials while keeping her total bid at most one.
We first use a partial allocation lemma of Seddighin and Seddighin to reduce the number of unserved agents from \({n}\) to at most \({n}/42\) within three layers. We then apply Theorem~\ref{thm:amplification} to the remaining agents and show that the number of unserved agents decreases sufficiently quickly to prove the desired bound.

\begin{lemma}[Seddighin--Seddighin~{\cite{SS25}}]\label{lem:ss-input}
Suppose $t$ agents have normalized, monotone, subadditive valuations on a finite ground set $M$, with $\MMS_t(v_i;M)=1$ for every agent. For every nonempty specified subset $Q$ of the agents, let $k=\lfloor t/|Q|\rfloor$. There is a feasible partial allocation giving value at least $1/2$ to at least $k|Q|/(k+1)$ agents in $Q$.
\end{lemma}

We use Lemma~\ref{lem:ss-input} only to construct the first three layers.
The following corollary extends its conclusion to the setting in which \(u\le {n}\) agents each have \({n}\) pairwise disjoint sets of value at least one.
The proof introduces \({n}-u\) auxiliary agents so that Lemma~\ref{lem:ss-input} can be applied with \({n}\) agents, while the allocation guarantee is required only for the original \(u\) agents.

\begin{corollary}\label{cor:partial}
Let $1\le u\le {n}$. Suppose \(u\) agents have normalized, monotone, subadditive valuations on \(M\), and each agent \(i\in[u]\) has \({n}\) pairwise disjoint sets \(W_{i,1},\ldots,W_{i,{n}}\) satisfying \(v_i(W_{i,r})\ge1\) for every \(r\in[{n}]\). Then there exists a feasible partial allocation giving value at least \(1/2\) to at least \(uk/(k+1)\) agents, where \(k=\lfloor {n}/u\rfloor\).
\end{corollary}

\begin{proof}
We construct an auxiliary instance with \({n}\) agents by adding \({n}-u\) copies of agent \(1\). Each copy is treated as a distinct agent with the same valuation as agent \(1\). Consequently, every agent in the auxiliary instance has \({n}\) pairwise disjoint sets of value at least one.

For each agent \(i\) in the auxiliary instance, extend her \(n\) disjoint sets to a partition of \(M\) into \(n\) bundles by assigning the remaining goods arbitrarily.
By monotonicity, every bundle in this partition has value at least one. Therefore, the auxiliary MMS benchmark \(\lambda_i=\MMS_n(v_i;M)\) satisfies \(\lambda_i\ge1\) for every agent $i$.

We divide each agent's valuation by her auxiliary MMS benchmark \(\lambda_i\). Since \(\lambda_i>0\), the resulting valuation remains normalized, monotone, and subadditive, and its MMS benchmark with respect to \({n}\) agents is exactly one.
Thus, Lemma~\ref{lem:ss-input} applies with \(t={n}\) and \(Q\) consisting of the \(u\) original agents.
Since \(k=\lfloor {n}/u\rfloor\), the lemma guarantees a feasible partial allocation giving value at least \(1/2\), under the scaled valuations, to at least \(uk/(k+1)\) agents in \(Q\).

Returning to the original valuations, each of the agents served by Lemma~\ref{lem:ss-input} receives a bundle \(A_i\) satisfying \(v_i(A_i)\ge\lambda_i/2\ge1/2\).
If the allocation assigns any bundles to the auxiliary agents, we discard those bundles, leaving their goods unassigned.
This does not affect the bundles or values of the original agents, and the resulting partial allocation remains feasible.
Hence, at least \(uk/(k+1)\) of the original \(u\) agents receive value at least \(1/2\), as required.
\end{proof}

\begin{proposition}[Few allocation layers]\label{prop:layers}
Suppose \({n}\ge2\) agents have normalized, monotone, subadditive valuations on \(M\), and each agent \(i\in[{n}]\) has \({n}\) pairwise disjoint sets \(W_{i,1},\ldots,W_{i,{n}}\subseteq M\) satisfying \(v_i(W_{i,r})\ge1\) for every \(r\in[{n}]\). Then there exists a multi-allocation giving every agent value at least \(1/2\), with multiplicity at most \(1+\log_2^*{n}\).
\end{proposition}

\begin{proof}

Let \(u_0={n}\), and for each \(j\ge1\), let \(u_j\) denote the number of unserved agents after the \(j\)-th layer, with \(u_j=0\) if the construction has already terminated.
The construction terminates as soon as all agents have been served.
We first apply Corollary~\ref{cor:partial} to construct at most three initial layers and then use Theorem~\ref{thm:amplification} to construct the subsequent layers. If exactly one agent remains unserved at any stage, we can serve her in one additional layer by assigning her any one of her \({n}\) sets of value at least one.
Since each layer is feasible and gives every agent it serves value at least $1/2$, it suffices to show that the construction terminates after at most $1+\log_2^*n$ layers.

\paragraph{The first three layers.}
We first show that the number of unserved agents after the first three layers is at most \({n}/42\). Suppose \(u_{j-1}>0\) agents remain unserved at the beginning of layer \(j\), where \(j\in\{1,2,3\}\).
Since every unserved agent still has \({n}\) pairwise disjoint sets of value at least one, Corollary~\ref{cor:partial} guarantees a feasible partial allocation in which at least \(u_{j-1}k/(k+1)\) agents receive value at least \(1/2\), where \(k=\lfloor {n}/u_{j-1}\rfloor\). We record the bundles of these agents and exclude them from subsequent layers. Consequently, the number of unserved agents after layer \(j\) satisfies
\[
u_j\le
\left\lfloor
\frac{u_{j-1}}{\lfloor {n}/u_{j-1}\rfloor+1}
\right\rfloor.
\]

Initially, \(u_0={n}\), so the first layer leaves at most \(\lfloor {n}/2\rfloor\) agents unserved. If \(u_1>0\), then \(u_1\le {n}/2\), which implies \(\lfloor {n}/u_1\rfloor\ge2\). Applying the preceding inequality to the second layer gives \(u_2\le\lfloor u_1/3\rfloor\le\lfloor {n}/6\rfloor\). Similarly, if \(u_2>0\), then \(u_2\le {n}/6\), so \(\lfloor {n}/u_2\rfloor\ge6\). The third layer therefore leaves at most \(u_3\le\lfloor u_2/7\rfloor\le\lfloor {n}/42\rfloor\) agents unserved. If the construction terminates before the third layer, the same inequalities hold because the subsequent values of \(u_j\) are defined to be zero. Thus,
\[
u_1\le\left\lfloor\frac {n}2\right\rfloor,\qquad
u_2\le\left\lfloor\frac {n}6\right\rfloor,\qquad
u_3\le\left\lfloor\frac {n}{42}\right\rfloor.
\]
In particular, after at most three layers, either all agents have been served or \(1\le u_3\le {n}/42\). In the latter case, every unserved agent still has \({n}\) pairwise disjoint sets of value at least one, while the number of unserved agents has decreased to at most \({n}/42\).
We next use Theorem~\ref{thm:amplification} to bound the number of agents that remain unserved after each subsequent layer.

\paragraph{Amplified layers.}
Suppose that \(u_3>0\), so \(u_3\le {n}/42\). Consider any \(j\ge3\) such that \(u_j\ge2\).
Since an agent who has been served is excluded from all subsequent layers, the number of unserved agents cannot increase. Therefore, \(u_j\le {n}/42\), or equivalently, \({n}/u_j\ge42\).
To construct the next layer, we apply Theorem~\ref{thm:amplification} to the \(u_j\) unserved agents, each of whom has \({n}\) pairwise disjoint sets of value at least one.
Choose the number of trials to be \(\ell_j=\lfloor {n}/[4(u_j-1)]\rfloor\). Since \(u_j\ge2\) and \({n}/u_j\ge42\), we have \(\ell_j\ge1\). Moreover, \(4\ell_j(u_j-1)\le {n}\), so the second conclusion of Theorem~\ref{thm:amplification} applies.

We first give a lower bound on \(\ell_j\). Using \(\lfloor x\rfloor\ge x-1\) for every real number \(x\), together with \(u_j-1\le u_j\), we obtain
\[
\ell_j
\ge\frac{{n}}{4(u_j-1)}-1
\ge\frac{{n}}{4u_j}-1
\ge\frac{5{n}}{24u_j},
\]
where the last inequality follows from \({n}/u_j\ge42\ge24\).

By Theorem~\ref{thm:amplification}, there exists a feasible partial allocation of the \(u_j\) unserved agents in which at most \(\lfloor u_j(3/4)^{\ell_j}\rfloor\) agents receive value below \(1/2\).
We choose such an allocation for layer \(j+1\), record the bundles of agents receiving value at least \(1/2\), and exclude these agents from subsequent layers.
Consequently, the number of unserved agents after this layer satisfies \(u_{j+1}\le u_j(3/4)^{\ell_j}\).

To express this upper bound in terms of \({n}/u_j\), observe that \((4/3)^5>4\), which implies \(\log_2(4/3)>2/5\). Combining this inequality with the preceding lower bound on \(\ell_j\), we obtain
\begin{equation}\label{eq:residual-recurrence}
u_{j+1}
\le u_j\left(\frac34\right)^{\ell_j}=u_j2^{-\ell_j\log_2(4/3)}\le u_j2^{-{n}/(12u_j)}.
\end{equation}
Thus, whenever \(u_j\ge2\), we can construct the next layer so that \eqref{eq:residual-recurrence} holds. In particular, \(u_{j+1}<u_j\), and hence the number of unserved agents strictly decreases. If \(u_j=1\), we instead assign the remaining agent any one of her \({n}\) sets of value at least one, so that \(u_{j+1}=0\). Therefore, the construction can continue until every agent has been served.

\paragraph{Bounding the total number of layers.}
We now show that all agents can be served in at most \(1+\log_2^*{n}\) layers. Recall from~\eqref{eq:towers} that \(E_0=1\) and \(E_{r+1}=2^{E_r}\) for every integer \(r\ge0\), and that \(\log_2^*{n}\) is the smallest nonnegative integer \(r\) satisfying \({n}\le E_r\).

First, suppose that \(2\le {n}\le4\). The bounds for the initial layers give \(u_2\le\lfloor {n}/6\rfloor=0\). Thus, all agents are served within two layers, which is at most \(1+\log_2^*{n}\).
For the remainder of the proof, suppose \({n}>4\) and let \(k=\log_2^*{n}\). Then \(k\ge3\) and \({n}\le E_k\). If \(u_3=0\), the construction has already terminated within three layers, and the desired bound follows. We may therefore assume that \(u_3>0\).

For each \(j\ge3\) with \(u_j>0\), let \(z_j=n/(12u_j)\).
By the bound \(u_3\le {n}/42\), we have \(z_3\ge42/12>3\). Suppose that \(u_j\ge2\) and \(u_{j+1}>0\). Applying \eqref{eq:residual-recurrence} and taking reciprocals gives
\begin{equation}\label{eq:tower-growth}
z_{j+1}
=\frac{{n}}{12u_{j+1}}
\ge\frac{{n}}{12u_j}2^{{n}/(12u_j)}
=z_j2^{z_j}.
\end{equation}
We claim that, for every \(j\ge4\), if \(u_j>0\), then \(z_j\ge E_{j-1}\). We prove this claim by induction on \(j\).

Suppose first that \(u_4>0\). Since a layer beginning with exactly one unserved agent would serve that agent directly, we must have \(u_3\ge2\). Thus, \eqref{eq:tower-growth} applies. Using \(z_3>3\), we obtain
\[
z_4\ge z_3 2^{z_3}>3\cdot2^3=24>E_3.
\]
This proves the claim for \(j=4\).

Now suppose that the claim holds for some \(j\ge4\), and that \(u_{j+1}>0\). Since the construction would terminate in the next layer if \(u_j=1\), we necessarily have \(u_j\ge2\). By \eqref{eq:tower-growth} and the induction hypothesis, we obtain
\[
z_{j+1}
\ge z_j2^{z_j}
\ge2^{E_{j-1}}
=E_j,
\]
where the second inequality uses \(z_j\ge E_{j-1}\ge1\). This proves the claim for \(j+1\). Consequently, by induction, \(z_j\ge E_{j-1}\) whenever \(j\ge4\) and \(u_j>0\).

Finally, we show that the construction must terminate by the end of layer \(k+1\). Suppose, for contradiction, that \(u_{k+1}>0\). Since \(k\ge3\), the preceding induction applies with \(j=k+1\), giving
\[
z_{k+1}\ge E_k\ge {n}.
\]
On the other hand, \(u_{k+1}\) is a positive integer, so \(u_{k+1}\ge1\). By the definition of \(z_{k+1}\), we have \(z_{k+1}={n}/(12u_{k+1})\le {n}/12<{n}\), contradicting the preceding inequality. Therefore, \(u_{k+1}=0\), and all agents are served after at most \(k+1=1+\log_2^*{n}\) layers.

Every agent receives a recorded bundle of value at least $1/2$, and each good belongs to at most one recorded bundle from each layer. Since the construction uses at most \(1+\log_2^*{n}\) layers, the resulting multi-allocation has multiplicity at most \(1+\log_2^*{n}\), as required.
\end{proof}

Corollary~\ref{cor:partial} is used only to reduce the number of unserved agents to at most \({n}/42\) within the first three layers.
If applied repeatedly, its guarantee would increase the ratio \({n}/u_j\) at least quadratically whenever agents remain unserved.
In contrast, the subsequent layers constructed using Theorem~
\ref{thm:amplification} satisfy the stronger recurrence \(z_{j+1}\ge z_j2^{z_j}\), where \(z_j={n}/(12u_j)\).
This exponential growth yields the iterated-logarithmic bound on the number of layers.
Appendix~\ref{app:bidding} provides an alternative initialization using Theorem~\ref{thm:amplification} without Corollary~\ref{cor:partial}, giving a multi-allocation with multiplicity at most \(19+\log_2^*{n}\).
This provides another derivation of the iterated-logarithmic bound, while the result of Seddighin and Seddighin allows us to obtain the sharper bound in Proposition~\ref{prop:layers}.

\subsection{Completing the subadditive allocation}
We now combine Proposition~\ref{prop:layers} and Feige's rounding theorem to prove Theorem~\ref{thm:sa-main}. Let $D=D(n)$ and $\alpha=\alpha(n)=1/(4D-2)$.
Since $\alpha$ is nonincreasing in $n$, the singleton reduction allows us to work with $n\ge2$ agents whose MMS values are one and whose singleton values are below $\alpha$. Each agent $i$ has the fixed MMS partition $(W_{i,1},\ldots,W_{i,n})\in\Pi(M,n)$.

The layer construction gives bundles of value at least $1/2$ that may overlap. The singleton bound controls the loss when we make these bundles disjoint by rounding.
We summarize the construction on the working instance in Algorithm~\ref{alg:subadditive}.

\begin{algorithm}[h]
\caption{Subadditive allocation by amplification and rounding}\label{alg:subadditive}
\begin{algorithmic}[1]
\Require Working instance with $n\ge2$ agents, normalized, monotone, subadditive valuations $v_i$ with $\MMS_n(v_i;M)=1$ and singleton values below $\alpha=\alpha(n)$, and MMS partitions $(W_{i,1},\ldots,W_{i,n})\in\Pi(M,n)$ for every $i\in[n]$.
\State Let $D=D(n)=2^{\lceil\log_2(1+\log_2^*n)\rceil}$, $\alpha=1/(4D-2)$, and $V=[n]$. Let $B_i=\varnothing$ for every $i\in[n]$.
\For{$j=1,2,3$}
\State If \(V=\varnothing\), exit the loop.
\State Let $u=|V|$ and $k=\lfloor n/u\rfloor$. Enumerate feasible partial allocations $(C_i)_{i\in V}$ on $M$ until at least $uk/(k+1)$ agents $i\in V$ satisfy $v_i(C_i)\ge1/2$, as guaranteed by Corollary~\ref{cor:partial}.
\State For every such agent $i$, let $B_i=C_i$ and remove $i$ from $V$. All goods remain available.
\EndFor
\While{\(V\ne\varnothing\)}
\State If \(|V|=1\), let \(i\) be the remaining agent, set \(B_i=W_{i,1}\), remove \(i\) from \(V\), and exit the loop.
\State Let $u=|V|$ and $\ell=\lfloor n/[4(u-1)]\rfloor$. For every $i\in V$ and $r\in[n]$, construct $\sigma_{i,r}=\textsc{BidDistribution}(W_{i,r},v_i,1/\ell)$.
\State Enumerate the finitely many realizations of $\textsc{RepeatedBidding}$ with agents $V$, sets $W_{i,r}$, distributions $\sigma_{i,r}$, and $\ell$ trials until an outcome $(C_i)_{i\in V}$ leaves at most $\lfloor u(3/4)^\ell\rfloor$ agents below value $1/2$, as guaranteed by Theorem~\ref{thm:amplification}.

\State For every $i\in V$ satisfying $v_i(C_i)\ge1/2$, let $B_i=C_i$ and remove $i$ from $V$. All goods remain available.
\EndWhile
\State For every $i\in[n]$, let $\delta_i=\max_{g\in B_i}v_i(\{g\})$.
\State Enumerate pairwise disjoint sets $A_i\subseteq B_i$, for $i\in[n]$, until, for every $i\in[n]$,
\[
Dv_i(A_i)\ge v_i(B_i)-(D-1)\delta_i,
\]
as guaranteed by Theorem~\ref{thm:rounding}.
\State Distribute all unassigned goods arbitrarily and return $(A_i)_{i\in[n]}$.
\end{algorithmic}
\end{algorithm}

\begin{proof}[Proof of \cref{thm:sa-main}]
By the singleton reduction, it suffices to verify that Algorithm~\ref{alg:subadditive} gives every agent value at least $\alpha$ on the working instance.
Every agent has $n$ pairwise disjoint MMS bundles of value at least one. The layer construction in Proposition~\ref{prop:layers} therefore gives bundles $(B_i)_{i\in[n]}$ satisfying $v_i(B_i)\ge1/2$ for every agent, with multiplicity at most
\[
1+\log_2^*n=d(n)\le D.
\]
Thus $(B_i)_{i\in[n]}$ is a $D$-multi-allocation. The chosen $D$ is a power of two and is at least two because $n\ge2$.

Since every singleton has value below $\alpha$, the largest singleton value $\delta_i$ in each $B_i$ is below $\alpha$. By Theorem~\ref{thm:rounding}, there are pairwise disjoint bundles $A_i\subseteq B_i$ such that, for every $i\in[n]$,
\[
v_i(A_i)\ge\frac{v_i(B_i)}{D}-\frac{D-1}{D}\delta_i
\ge\frac{1}{2D}-\frac{D-1}{D}\alpha=\alpha,
\]
where the equality follows from $(2D-1)\alpha=1/2$. Every agent's MMS is one, so these bundles meet the required MMS guarantee. Distributing unassigned goods gives a complete allocation without decreasing any agent's value.

Finally, since \(D(n)\) is the smallest power of two at least \(d(n)=1+\log_2^*n\), we have \(D(n)<2d(n)\). Therefore,
\[
\alpha(n)
=\frac{1}{4D(n)-2}
>
\frac{1}{8(1+\log_2^*n)}.
\]
This proves Theorem~\ref{thm:sa-main}.
\end{proof}

\section{MMS allocations in the partitioning hierarchy}\label{sec:partitioning}
We prove \cref{thm:p-main} by constructing allocations through improving transfers.
If an agent is below the current target, we consider a whole MMS bundle or, in the second construction, one of its two complementary pieces as a \emph{candidate bundle} to transfer to her: we add its goods to her bundle and remove them from the other participating agents' bundles. For a fixed recipient, we call the other participating agents \emph{donors}. We compare the recipient's gains with the donors' losses in a bounded potential, summing over a selected disjoint family of candidate bundles. We evaluate each transfer separately from the same current allocation. Supporting prices allow us to control these losses across disjoint candidate bundles. If the total gain exceeds the total loss, at least one transfer increases the potential. We carry out one such transfer and repeat.
In each construction, the potential is bounded, and every iteration increases it by at least a fixed positive amount. Therefore, the procedure terminates with every participating agent meeting the target.

We apply this argument in two ways. In Section~\ref{sec:grouping}, we use \emph{capped social welfare}, the sum of the bundle values after each is capped at one, and whole MMS bundles as candidates. We divide the agents into groups to reduce the number of donors, construct a feasible allocation for each group using the same goods, and then round the resulting multi-allocation. In Section~\ref{sec:direct}, we keep all agents together and use both whole MMS bundles and their complementary pieces. A different potential makes these additional candidates useful while keeping each donor's total loss bounded, giving a direct construction that maintains feasibility throughout. We first prove the loss bounds from supporting prices and explain why the direct comparison with capped social welfare gives only a weak guarantee. We combine the guarantees of the two constructions in Section~\ref{sec:combining} to prove \cref{thm:p-main}.

\subsection{Supporting prices and disjoint deletion losses}
In this section, we bound the sum of the losses caused by separately deleting disjoint subsets of a bundle. To apply it to capped social welfare, we first show that capping preserves the supporting-price property. For a valuation $v$ and $C\ge0$, we define its capped valuation by $v^C(S)=\min\{v(S),C\}$ for every $S\subseteq M$. Capping at $C$ then turns a bound proportional to a donor's capped value into a bound proportional to $C$.

\begin{lemma}\label{lem:closure}
The class $\Q_q$ is closed under restriction to a subset of goods, nonnegative scaling, and capping at a constant: if $v\in\Q_q$ and $C\ge0$, then $v^C$ belongs to $\Q_q$.
\end{lemma}
\begin{proof}
Restricting the ground set to a subset $R\subseteq M$ does not change the valuation of any set $S\subseteq R$ or the supporting prices available for its partitions.
Multiplying the valuation by a nonnegative constant allows us to multiply all supporting prices by the same constant. Both operations therefore preserve~\eqref{eq:group-prices}.

To prove closure under capping, we fix a set $S$ and a partition of $S$ into at most $q$ groups. Suppose first that $v(S)>0$. We take supporting prices $w_1,\ldots,w_t$ for this partition, let $\theta=\min\{1,C/v(S)\}$, and multiply every price by $\theta$. The new prices sum to $\theta v(S)=v^C(S)$. To check a subtotal, we let $U$ be the union of any chosen groups. The corresponding new subtotal is at most $v(U)$ because $\theta\le1$ and the original prices satisfy the supporting-price inequalities. It is also at most the total new price, which is at most $C$, because all prices are nonnegative. Thus it is at most $\min\{v(U),C\}=v^C(U)$, as required. If $v(S)=0$, monotonicity gives value zero on every subset of $S$, and it suffices to use zero prices.
\end{proof}

We next use supporting prices to bound the sum of value losses caused by deleting pairwise disjoint subsets separately from the same bundle.

\begin{lemma}\label{lem:deletion}
Let $L\ge0$ be an integer, let $f\in\Q_q$, and let $B_1,\ldots,B_L$ be pairwise disjoint subsets of a set $S\subseteq M$. Then
\[
 \sum_{\ell=1}^{L}\bigl(f(S)-f(S\setminus B_\ell)\bigr)
 \le\max\{1,L/q\}\,f(S).
\]
\end{lemma}
\begin{proof}
For $L=0$, the claim is immediate. Since empty deletion sets contribute zero loss and the right-hand side is nondecreasing in $L$, we may omit them and reindex the remaining sets. If $1\le L\le q$, we enlarge one of the sets by the goods outside their union to obtain a partition of $S$ into $L$ groups. Let $w_1,\ldots,w_L$ be supporting prices for this partition. Deleting the $\ell$th group leaves the union of all other groups, whose value is at least $f(S)-w_\ell$. Deleting only the original $B_\ell$ leaves at least as much value. Hence
\[
 f(S)-f(S\setminus B_\ell)\le w_\ell,
 \qquad
 \sum_{\ell=1}^L w_\ell=f(S).
\]
This proves the claim for $L\le q$.

For $L>q$, consider the \(q\) sets with the largest deletion losses.
By the preceding argument, their total loss is at most \(f(S)\).
Every other deletion loss is at most the average of these \(q\) losses.
Therefore, the sum of all \(L\) deletion losses is at most \((L/q)f(S)\).
\end{proof}

To apply Lemma~\ref{lem:deletion} to
bundle transfers, we consider a disjoint family of candidate bundles.
For each donor, the intersections of these candidates with her current bundle are pairwise disjoint and represent the goods she would lose in the corresponding transfers.
We use Lemma~\ref{lem:deletion} to bound the sum of these losses under capped social welfare.
For the potential in Section~\ref{sec:direct}, we will use supporting prices together with a scalar criterion to control the corresponding losses.

\paragraph{The direct comparison with capped social welfare.}
Consider an $n$-agent instance with $\mu_i=1$ for every agent $i$. Even when $q\ge n$, the direct comparison using whole MMS bundles and capped social welfare does not yield a constant MMS guarantee.
Fix an agent \(i\) whose current capped value is \(z\).
For each of her \(n\) MMS bundles, consider transferring that bundle to agent \(i\), with every transfer evaluated separately from the same current allocation.
Each transfer increases her capped value from \(z\) to one, so the sum of her gains over these \(n\) candidate transfers is \(n(1-z)\).
By \cref{lem:closure,lem:deletion}, each of the other $n-1$ agents loses at most one in total over these transfers.
Therefore, the sum of the changes in capped social welfare is at least
\[
 n(1-z)-(n-1)=1-nz.
\]
This lower bound implies that an improving transfer exists when $z<1/n$, but gives no constant target as $n$ grows. The term $nz$ arises because we subtract the recipient's current capped value once for every candidate. We can improve the comparison in two ways. In Section~\ref{sec:grouping}, we use the same candidates and capped social welfare but reduce the number of donors. In Section~\ref{sec:direct}, we use a potential whose term for a recipient below the target is zero, so no current contribution is subtracted, and use pieces of MMS bundles to increase her total gain.

\subsection{Grouping and rounding}\label{sec:grouping}
We first divide agents into groups to reduce the number of donors in each transfer, while keeping all $n$ MMS bundles of each recipient as candidates and using capped social welfare as the potential.
Within each group, only the other group members can lose value from a transfer.
We construct a feasible partial allocation for each group on the same set of goods.
The resulting bundles may overlap across groups, and we apply Feige's rounding theorem to make them disjoint.
Increasing the number of groups improves the intermediate value guarantee but also increases the rounding loss.
We choose the number of groups to balance these two effects.
For example, when $q\ge n$, two groups give every agent value at least $1/2$ before rounding and at least $1/6$ afterward.
Our construction is summarized in Algorithm~\ref{alg:grouping}.

\begin{algorithm}

\caption{Grouping and rounding}\label{alg:grouping}
\begin{algorithmic}[1]
\Require Working instance with integers $n,q\ge2$, valuations $v_i\in\Q_q$ with $\mu_i=1$, and
MMS partitions with bundle collections $\mathcal P_i=\{W_{i,1},\ldots,W_{i,n}\}$, and $v_i(\{g\})<\alpha$ for all $i,g$, where $\alpha$ is defined in Line~\ref{step::alg2-1}.
\State Set $x=\max\{1,n/q\}$ and $\alpha=1/(11x-5)$. Choose the smallest power of two $d\ge2$ with $x\le4d^2/(6d-1)$ and set $\tau=1-x/d$.\label{step::alg2-1}
\State Set $B_i=\varnothing$ for every $i\in[n]$ and partition $[n]$ into $d$ groups whose sizes differ by at most one, allowing empty groups.
\For{each nonempty group $G$}
 \While{some $i\in G$ has $v_i(B_i)<\tau$}
  \State Choose an agent $i\in G$ with $v_i(B_i)<\tau$.
  \State For each $W\in\mathcal P_i$, set $B_i^W=B_i\cup W$ and $B_j^W=B_j\setminus W$ for every $j\in G\setminus\{i\}$.
  \State Choose $W \in {\mathcal P_i}$ maximizing $\sum_{j\in G}\min\{v_j(B_j^W),1\}$.
  \State Set $B_j=B_j^W$ for every $j\in G$.
 \EndWhile
\EndFor
\State For each $i\in[n]$, set $\delta_i=\max_{g\in B_i}v_i(\{g\})$, with $\max\varnothing=0$.
\State Enumerate pairwise disjoint $A_i\subseteq B_i$ until $d\,v_i(A_i)\ge v_i(B_i)-(d-1)\delta_i$ for every $i\in[n]$, as guaranteed by Theorem~\ref{thm:rounding}.
\State \Return the bundles $(A_i)_{i\in [n]}$, distributing unassigned goods arbitrarily.
\end{algorithmic}
\end{algorithm}

\begin{proposition}[Guarantee from grouping agents]\label{prop:grouping}
Let $n,q\ge2$ be integers, $x=\max\{1,n/q\}$, and $\alpha=1/(11x-5)$. Algorithm~\ref{alg:grouping} returns an $\alpha$-MMS allocation on the working instance. Consequently, every instance with valuations in $\Q_q$ admits an $\alpha$-MMS allocation.
\end{proposition}
\begin{proof}

We analyze Algorithm~\ref{alg:grouping} on the working instance. Let \(d\) be the power of two chosen in Line~\ref{step::alg2-1}.
Since \(x\le 4d^2/(6d-1)<d\), the intermediate target \(\tau=1-x/d\) is positive.

\paragraph{The gain from grouping.}
We partition the agents into $d$ groups whose sizes differ by at most one, allowing empty groups. To analyze one group, we fix a nonempty group $G$ of size $s$. Its number of donors satisfies $s-1<n/d$, so $n-d(s-1)\ge1$. Let $m_0=\min\{q,n\}$, so $n/m_0=x$.

We start with empty bundles for this group and use the capped social welfare
\[
 \Phi_G=\sum_{j\in G}\min\{v_j(B_j),1\}.
\]
If agent $i\in G$ has capped value $z$, transferring any of her $n$ MMS bundles to her raises it to one. Her total gain over these transfers is $n(1-z)$. By \cref{lem:closure,lem:deletion}, each donor's total loss is at most $\max\{1,n/q\}=n/m_0$. Dividing the resulting total welfare change by $n$, the average increase is at least
\[
 1-z-\frac{s-1}{m_0}.
\]
Because $(s-1)/m_0<x/d$, this comparison suggests the intermediate target $\tau=1-x/d>0$. Whenever $z<\tau$, the average increase satisfies
\begin{equation*}
 1-z-\frac{s-1}{m_0}
 >\frac xd-\frac{s-1}{m_0}
 =\frac{n-d(s-1)}{dm_0}
 \ge\frac1{dm_0}.
\end{equation*}
We choose a transfer with the largest welfare increase. Since $0\le\Phi_G\le s$, fewer than $sdm_0$ such transfers can occur. If a group member were still below $\tau$, the same comparison would give another increase. Thus our procedure ends with every member meeting $\tau$.

\paragraph{The cost of rounding.}
We construct each group allocation on the same set of goods $M$. A good is used at most once within a group and hence at most $d$ times across groups. The resulting $d$-multi-allocation gives every agent value at least $\tau$. The valuations are subadditive, $d$ is dyadic, and every singleton is worth less than $\alpha$. We can therefore apply \cref{thm:rounding} to obtain disjoint subsets of the group bundles giving every agent value at least
\[
 \frac{\tau-(d-1)\alpha}{d}.
\]
To make this value at least $\alpha$, it suffices to have $\alpha\le\tau/(2d-1)$. We denote this upper bound on the admissible target by $F_d(x)$:
\begin{equation}\label{eq:Fd}
 F_d(x)=\frac{(1-x/d)_+}{2d-1}.
\end{equation}
Here $t_+=\max\{t,0\}$ extends the formula to $d\le x$, when the bound is zero. The numerator records the value reached within each group. The denominator accounts for the division by $d$ and the additional singleton loss in rounding.

\paragraph{Choosing the number of groups.}
By \cref{lem:envelope} (in Appendix~\ref{app:envelope}), the value of $d$ chosen in Algorithm~\ref{alg:grouping} satisfies $F_d(x)\ge1/(11x-5)=\alpha$.
Thus the rounded allocation gives every agent value at least $\alpha$, proving the proposition.
\end{proof}

When $q\ge n$, we have $x=1$ and $d=2$, so $\tau=1/2$ and $\alpha=1/6$. For $q\le n$, the approximation factor is $q/(11n-5q)$.

\subsection{Direct transfers from whole MMS bundles and pieces}\label{sec:direct}\label{sec:direct-potential}
For $q>n$, we apply the transfer argument to all agents together, maintaining a feasible allocation throughout.
To obtain an improving transfer, we modify the potential and increase the recipient's gain by considering additional candidate bundles.
We use a potential whose term for an agent below the target is zero. For each MMS bundle $W$, we choose a subset $B\subseteq W$ and consider transferring the whole bundle $W$ or either of the two complementary pieces $B$ and $W\setminus B$.
Replacing a whole bundle by its two pieces adds one candidate to a disjoint comparison family, so a larger $q$ allows us to use more splits.
We choose the potential so that splitting increases the recipient's total gain while keeping each donor's total loss bounded.

\paragraph{Choosing the potential.}

We use the potential $\Phi=\sum_{i=1}^n\psi_{\alpha}(v_i(A_i))$, where $\psi_{\alpha}$ is a function to be chosen. In order to obtain an improving transfer, we require three properties:
\begin{itemize}
\item  First, $\psi_{\alpha}$ is nondecreasing and zero on $[0,\alpha]$, so a recipient below the target gains at least $\psi_{\alpha}(v_i(B))$ when she receives a candidate bundle $B$.
\item Second, for transfers of at most $q$ disjoint candidate bundles, the sum of each donor's losses in the potential is at most one, with every transfer evaluated from the same allocation.
\item Third, we can select a random disjoint family of at most $q$ whole MMS bundles or pieces whose transfers give the recipient expected total gain at least $n$.
\end{itemize}
These properties give a positive lower bound on the expected sum of the potential changes:
\[
 \underbrace{n}_{\text{recipient's expected total gain}}
 -\underbrace{(n-1)}_{\text{donors' total loss}}>0.
\]
We first construct \(\psi_\alpha\) and bound the recipient's gains from whole MMS bundles and their pieces.
We then use a randomly selected disjoint family of candidates to prove the existence of an improving transfer.

\paragraph{Contributions from a whole bundle and its pieces.}
Fix $0<\alpha\le1/4$ and let $\beta=(1-\alpha)/2$. Consider an agent with valuation $v$ and an MMS bundle $W$ of value at least one whose singletons are worth less than $\alpha$. To split $W$, we start with the empty set and add goods of $W$ until their value first reaches $\beta$. We call the resulting set $B$. By subadditivity and the value of the last singleton, we have
\[
 \beta\le v(B)<\beta+\alpha,
\]
and
\[
v(W\setminus B)\ge1-v(B).
\]
The identity $2\beta+\alpha=1$ puts both $v(B)$ and $1-v(B)$ in $[\beta,\beta+\alpha]$. If we choose $\psi_{\alpha}$ to be linear on this interval, then $\psi_{\alpha}(t)+\psi_{\alpha}(1-t)$ is constant there. We will use this property to obtain a fixed lower bound on the combined contribution of the two pieces.

The donor-loss argument will require $2\psi_\alpha(x)-\psi_\alpha(x-\alpha)\le1$ for $x\ge2\alpha$. Where the left-hand side is constant and both derivatives exist, $2\psi_{\alpha}'(x)=\psi_{\alpha}'(x-\alpha)$. Thus, in order to keep this bound tight, we are led to successive slope reductions by a factor of two. We start the function at zero at $\alpha$ with slope $1/\beta$, halve its slope at $\beta$ and again at $\beta+\alpha$, and make it constant at $\beta+2\alpha$. Explicitly,
\begin{equation}\label{eq:middle-potential}
 \psi_{\alpha}(x)=
 \begin{cases}
  0,&0\le x\le \alpha,\\[2pt]
  (x-\alpha)/\beta,&\alpha\le x\le \beta,\\[2pt]
  (x+\beta-2\alpha)/(2\beta),&\beta\le x\le \beta+\alpha,\\[2pt]
  (x+3\beta-3\alpha)/(4\beta),&\beta+\alpha\le x\le \beta+2\alpha,\\[2pt]
  1-\alpha/(4\beta),&x\ge \beta+2\alpha.
 \end{cases}
\end{equation}
This function is continuous, nondecreasing on $[0,\infty)$, and concave on $[\alpha,\infty)$.
Let $H$ denote its maximum. We denote by $C$ the common value of $\psi_\alpha(t)+\psi_\alpha(1-t)$ for $t\in[\beta,\beta+\alpha]$. Then
\begin{equation}\label{eq:middle-contributions}
 \begin{aligned}
 H&=1-\frac{\alpha}{4\beta}<1,\\
 C&\mathrel{=}2-\frac{3\alpha}{2\beta}.
 \end{aligned}
\end{equation}
Since $\beta+2\alpha<1$, a whole MMS bundle contributes $H$. For the two pieces, monotonicity and $v(W\setminus B)\ge1-v(B)$ give
\[
 \psi_\alpha(v(B))+\psi_\alpha(v(W\setminus B))
 \ge\psi_\alpha(v(B))+\psi_\alpha(1-v(B))=C.
\]
The equality uses $v(B),1-v(B)\in[\beta,\beta+\alpha]$; the actual value $v(W\setminus B)$ may be larger than $\beta+\alpha$. For $\alpha\le1/4$, we have $C\ge1>H$, so splitting increases the guaranteed combined contribution.

The function also satisfies the required loss bound: for every $v\in\Q_q$ and pairwise disjoint $B_1,\ldots,B_L\subseteq S$ with $L\le q$,
\begin{equation}\label{eq:middle-deletion}
 \sum_{\ell=1}^L\bigl[\psi_{\alpha}(v(S))-\psi_{\alpha}(v(S\setminus B_\ell))\bigr]\le1.
\end{equation}
The resulting allocation procedure is presented in Algorithm~\ref{alg:direct}.

\begin{algorithm}[H]

\caption{Direct transfers from whole MMS bundles and pieces}\label{alg:direct}
\begin{algorithmic}[1]
\Require Working instance with integers $q>n\ge2$, valuations $v_i\in\Q_q$ with $\mu_i=1$, MMS partitions with bundle collections $\mathcal P_i=\{W_{i,1},\ldots,W_{i,n}\}$, and $v_i(\{g\})<\alpha$ for all $i,g$, where $\alpha$ is defined in Line~\ref{step:alg3-1}.
\State Set $r=\min\{q,2n\}$, $\alpha=2(r-n)/(7r-6n)$, $\beta=(1-\alpha)/2$, and use $\psi_{\alpha}$ from~\eqref{eq:middle-potential}.\label{step:alg3-1}
\For{each $i\in[n]$}
 \State Set $\mathcal C_i=\varnothing$ and $A_i=\varnothing$.
 \For{each $W\in\mathcal P_i$}
  \State Starting with $B=\varnothing$, add goods of $W$ one at a time until $v_i(B)\ge \beta$. Add $W,B$, and $W\setminus B$ to $\mathcal C_i$.
 \EndFor
\EndFor
\While{some $i\in[n]$ has $v_i(A_i)<\alpha$}
 \State Choose an agent $i\in[n]$ with $v_i(A_i)<\alpha$.
 \State For each $B\in\mathcal C_i$, define $A_i^B=A_i\cup B$ and $A_j^B=A_j\setminus B$ for every $j\in[n]\setminus\{i\}$.
 \State Choose {$B\in \mathcal{C}_i$} maximizing $\sum_{j\in[n]}\psi_{\alpha}(v_j(A_j^B))$, and set $A_j=A_j^B$ for every $j\in[n]$.
\EndWhile
\State \Return the bundles $(A_i)_{i\in [n]}$, distributing unused goods arbitrarily.
\end{algorithmic}
\end{algorithm}

\begin{proposition}[Direct guarantee from whole MMS bundles and pieces]\label{prop:middle}\label{prop:direct}
Let $q>n\ge2$ be integers, and let
\[
 \alpha=\min\left\{\frac14,\frac{2(q-n)}{7q-6n}\right\}.
\]
Algorithm~\ref{alg:direct} returns an $\alpha$-MMS allocation on the working instance. Consequently, every instance with valuations in $\Q_q$ admits an $\alpha$-MMS allocation.
\end{proposition}

\begin{proof}
Let $r=\min\{q,2n\}$ and $p=(r-n)/n\in(0,1]$.
{The target in Algorithm~\ref{alg:direct} satisfies $\alpha=2p/(1+7p)$. We use $p$ as the probability of splitting each MMS bundle in the comparison below.}

\paragraph{Candidates and disjoint families.}
For each agent $i$ and each MMS bundle $W\in\mathcal P_i$, we construct the two complementary pieces $B$ and $W\setminus B$ and consider transfers of $W$, $B$, and $W\setminus B$.
These three candidates are not pairwise disjoint. To bound the donors' losses, we instead consider disjoint families that contain either the whole bundle $W$ or its two pieces.
\[
 \begin{array}{c|c}
 \text{Candidates evaluated by the algorithm}&\text{Selection used in the proof}\\[2pt]
 W,\ B,\ W\setminus B&\{W\}\ \text{or}\ \{B,W\setminus B\}.
 \end{array}
\]
The algorithm's three candidates are not jointly disjoint: $W$ contains both pieces. Either selection used in our proof is disjoint. Replacing the whole bundle by its two pieces adds one candidate to the selected family. We use the random selection below only to prove that some candidate transfer improves the potential.

We choose uniformly at random $r-n$ of the recipient's $n$ MMS bundles.
From each chosen bundle, we take its two pieces; from every other bundle, we take the whole bundle.
The resulting family $\mathcal F$ is disjoint and has exactly $n+(r-n)=r$ candidates. Each MMS bundle is split with probability $p$.

\paragraph{The balance that determines the target.}
Since each MMS bundle is split with probability \(p\), its expected contribution is at least $(1-p)H+pC$:
the whole bundle contributes $H$, while its two pieces contribute at least $C$ in total.
To obtain total expected contribution at least $n$, it suffices to make this expression at least one. Substituting~\eqref{eq:middle-contributions} gives
\[
 (1-p)H+pC
 =1+\frac{2p-\alpha(1+7p)}{2(1-\alpha)}.
\]
Setting the numerator of the second term to zero gives exactly the chosen target $\alpha=2p/(1+7p)$. Consequently,
\begin{equation}\label{eq:middle-balance}
 \E\!\left[\sum_{B\in\mathcal F}\psi_{\alpha}(v_i(B))\right]
 \ge n\bigl[(1-p)H+pC\bigr]=n.
\end{equation}
Every comparison family has $|\mathcal F|=r=(1+p)n$ candidates.

\paragraph{An improving transfer.}
{Consider an iteration of Algorithm~\ref{alg:direct} in which agent $i$ satisfies $v_i(A_i)<\alpha$.
Since $\psi_\alpha$ is zero on \([0,\alpha]\), her current contribution to the potential is zero.
For each candidate \(B\in\mathcal C_i\), let \(\Delta_B\Phi\) denote the change in the potential caused by transferring \(B\) to agent \(i\).
By monotonicity, the recipient's gain is at least \(\psi_\alpha(v_i(B))\).}

{For each donor \(j\ne i\), the sets \(A_j\cap B\), where \(B\in\mathcal F\), are pairwise disjoint. Since \(|\mathcal F|=r\le q\), inequality (\ref{eq:middle-deletion}) bounds the sum of this donor's losses over these transfers by one. Combining this bound with (\ref{eq:middle-balance}), we obtain}
\[
 \E\!\left[\sum_{B\in\mathcal F}\Delta_B\Phi\right]
 \ge n-(n-1)=1.
\]
Every selected family \(\mathcal F\) contains exactly \(r\) candidates from \(\mathcal C_i\).
Hence
\[
1\le
\mathbb E\left[\sum_{B\in\mathcal F}\Delta_B\Phi\right]
\le r\max_{B\in\mathcal C_i}\Delta_B\Phi.
\]
Thus, the transfer selected by Algorithm~\ref{alg:direct} increases the potential by at least $1/r$.

The procedure keeps all \(n\) agents, including those already meeting the target, since subsequent transfers may decrease their values.
Starting from empty bundles, the potential is initially zero and satisfies \(0\le\Phi\le nH<n\) throughout the construction.
Since every iteration increases the potential by at least \(1/r\), fewer than \(nr\) transfers can occur.
If any agent were still below \(\alpha\), the same argument would guarantee another increase. Therefore, the procedure terminates with every agent meeting the target.

When \(q\ge2n\), we have \(r=2n\) and \(\alpha=1/4\), giving the claimed \(1/4\)-MMS guarantee.
\end{proof}

On the working instance, given its MMS values and partitions, constructing all pieces uses $O(nm)$ value queries. Each iteration evaluates at most $3n$ candidates for at most $n$ agents, using $O(n^2)$ queries. Since there are fewer than $nr\le2n^2$ transfers, Algorithm~\ref{alg:direct} uses $O(nm+n^4)$ value queries. This bound concerns the stated inputs for the working instance; it does not include finding MMS values or partitions during the singleton reduction.

\paragraph{Verifying the donor-loss bound.}
It remains to prove~\eqref{eq:middle-deletion}.
Lemma~\ref{lem:deletion} bounds the summed value losses from disjoint deletions. The following scalar criterion converts such an aggregate bound into a bound on losses after applying a function of the value.
The scalar argument also applies to valuations beyond \(\mathcal Q_q\), provided that the summed deletion losses are at most the value of the original bundle.

\begin{lemma}[From aggregate value losses to potential losses]\label{lem:potential-deletion}
Let $\alpha>0$ and let $f:[0,\infty)\to[0,\infty)$ be nondecreasing, zero on $[0,\alpha]$, and concave on $[\alpha,\infty)$. Suppose that the first inequality below holds for every $\alpha<x\le2\alpha$ and the second for every $x\ge2\alpha$:
\begin{equation}\label{eq:scalar-conditions}
 \begin{aligned}
 f(x)\le1-\alpha/x,\quad
 2f(x)-f(x-\alpha)\le1.
 \end{aligned}
\end{equation}
Let $v:2^M\to\R_{\ge0}$ be normalized and monotone, and let $L\ge0$ be an integer. For any $S\subseteq M$ and pairwise disjoint subsets $B_1,\ldots,B_L\subseteq S$ satisfying
\[
 \sum_{\ell=1}^L\bigl[v(S)-v(S\setminus B_\ell)\bigr]\le v(S),
\]
we have
\[
 \sum_{\ell=1}^L\bigl[f(v(S))-f(v(S\setminus B_\ell))\bigr]\le1.
\]
In particular, the conclusion holds for $v\in\Q_q$ and $L\le q$.
\end{lemma}
\begin{proof}
Let $x=v(S)$ and $p_\ell=v(S)-v(S\setminus B_\ell)$. Monotonicity gives $0\le p_\ell\le x$, and the hypothesis gives $\sum_\ell p_\ell\le x$. Since $x-p_\ell=v(S\setminus B_\ell)$, it suffices to prove the scalar inequality
\[
 \sum_{\ell=1}^L\bigl[f(x)-f(x-p_\ell)\bigr]\le1
\]
for every $x\ge0$ and every finite family of nonnegative numbers $p_\ell$ whose sum is at most $x$.

All scalar losses vanish when $x\le\alpha$.
Otherwise let $\Delta(p)=f(x)-f(x-p)$ for $0\le p\le x$.
Concavity of $f$ above $\alpha$ makes $\Delta$ convex on $[0,x-\alpha]$; it vanishes at zero and is constant thereafter. For $\alpha<x\le2\alpha$, its chord bound and the first condition in~\eqref{eq:scalar-conditions} give
\[
 \Delta(p)\le\frac{p f(x)}{x-\alpha}\le\frac px.
\]
This also holds for $p\ge x-\alpha$, where $\Delta(p)=f(x)$. Summing over the $p_\ell$ proves the scalar inequality in this range.

For $x\ge2\alpha$, let $T=x-\alpha\ge\alpha$ and truncate each $p_\ell$ at $T$, preserving $\Delta(p_\ell)$.
Their sum remains at most $x=T+\alpha\le2T$.
By convexity, transferring mass from a smaller positive number to a larger one, without exceeding \(T\), does not decrease the sum of their losses.
Repeating this operation until one of the two numbers becomes zero or \(T\), we obtain a family with at most one number strictly between zero and \(T\).
If one number reaches $T$, the others sum to at most $\alpha$ and at most one of them is positive; if none reaches $T$, at most one number is positive. Since $\Delta$ is nondecreasing, the summed loss is at most
\[
 \Delta(T)+\Delta(\alpha)=2f(x)-f(x-\alpha)\le1,
\]
by the second condition in~\eqref{eq:scalar-conditions}. The two ranges cover every $x>\alpha$. Finally, for $v\in\Q_q$ and $L\le q$, Lemma~\ref{lem:deletion} supplies the aggregate hypothesis.
\end{proof}

\begin{proof}[Verification of~\eqref{eq:middle-deletion}]
Our function $\psi_\alpha$ is nondecreasing, zero on $[0,\alpha]$, and concave above $\alpha$, so it suffices to check~\eqref{eq:scalar-conditions}.

For the first condition, we distinguish the two possible orders of $\beta$ and $2\alpha$.
\begin{enumerate}
\item If $0<\alpha\le1/5$, then $\beta\ge2\alpha$. For every $\alpha<x\le2\alpha$, the first positive segment of $\psi_\alpha$ gives
\[
 \psi_\alpha(x)=\frac{x-\alpha}{\beta}
 \le\frac{x-\alpha}{x}=1-\frac\alpha x.
\]
\item If $1/5<\alpha\le1/4$, then $\beta<2\alpha$. The same comparison applies for $\alpha<x\le\beta$. On $[\beta,2\alpha]$, the function $\psi_\alpha$ is linear and satisfies
\[
 \psi_\alpha(\beta)=1-\frac\alpha\beta,
 \qquad
 \psi_\alpha(2\alpha)=\frac12=1-\frac\alpha{2\alpha}.
\]
It is therefore the chord of the concave function $1-\alpha/x$ between these endpoints, and lies below that function throughout the interval.
\end{enumerate}
These two cases prove the first condition for every $\alpha<x\le2\alpha$.

For the second condition, let $E(x)=2\psi_\alpha(x)-\psi_\alpha(x-\alpha)$ for $x\ge2\alpha$. Substituting the segments of~\eqref{eq:middle-potential} gives
\[
 E(x)=
 \begin{cases}
  x/\beta,&2\alpha\le x\le\beta\quad(\beta\ge2\alpha),\\[2pt]
  1,&\max\{2\alpha,\beta\}\le x\le\beta+2\alpha,\\[2pt]
  1-\dfrac{x-\beta-2\alpha}{4\beta},&\beta+2\alpha\le x\le\beta+3\alpha,\\[4pt]
  H,&x\ge\beta+3\alpha.
 \end{cases}
\]
Every displayed value is at most one. Thus the second condition holds for all $x\ge2\alpha$, and \cref{lem:potential-deletion} proves~\eqref{eq:middle-deletion}.
\end{proof}

\subsection[Combining the guarantees]{Combining the guarantees}\label{sec:combining}
The grouping construction applies for every $q\ge2$, while the direct construction applies for $q>n$. We complete the proof by comparing their guarantees where both apply and selecting the larger one.
\begin{proof}[Proof of \cref{thm:p-main}]
By the reduction in Section~\ref{sec:preliminaries}, we may assume $n\ge2$. Let $c=q/n$. If $c\le1$, we apply \cref{prop:grouping} to obtain $c/(11-5c)$. If $c\ge1$, the same proposition gives $1/6$.
For $1<c\le2$, \cref{prop:middle} gives $2(c-1)/(7c-6)$, which is at least $1/6$ exactly when $c\ge6/5$. For $c\ge2$, it gives $1/4$. These are the four branches of~\eqref{eq:rho}, and adjacent formulas agree at $c=1,6/5,2$.
\end{proof}

\section{A \texorpdfstring{$5/16$}{5/16}-MMS allocation for XOS valuations}
\label{sec:xos-mms}

To prove Theorem~\ref{thm:xos-main}, we use fixed additive lower
bounds of total value one on each agent's disjoint MMS bundles.
We construct a fractional allocation under auxiliary additive
valuations and round it to obtain disjoint bundles. Rounding may
lose the largest auxiliary coefficient, so we allow an agent to
receive selected goods integrally and use auxiliary coefficients
on the other goods. We call the selected goods her reserved goods.
We must choose these goods and the auxiliary coefficients together,
since different agents may need the same goods.

At a high level, we first allocate pairs that already meet an
agent's target and bound the resulting damage to the other agents'
MMS bundles. We then prove a general reservation theorem that
coordinates the agents' reserved goods and auxiliary valuations.
Finally, we construct these valuations and prove that they satisfy
the theorem's condition for the $5/16$ target. Algorithm~\ref{xos:alg:reservation} summarizes the reservation and rounding procedure.

\subsection{MMS bundles after allocating pairs}
\label{xos:subsec:reduction}

By the singleton reduction in Lemma~\ref{lem:singleton-reduction},
it suffices to consider the working instance from
Section~\ref{sec:preliminaries}. Let $\alpha=5/16$ be our target.
Every agent has MMS one, and every singleton has value less than
$\alpha$.

For each agent $i$ and each bundle $P$ in her MMS partition, we choose
supporting prices and divide them by \(v_i(P)\). Extended by zero
outside $P$, the resulting prices define a nonnegative additive lower bound $w_P$;
let $w_P(g)$ denote its coefficient on good $g$.
We call these numbers \emph{coefficients} since they may differ
from the agent's singleton values.
For every $S\subseteq M$, we have
\[
 w_P(P)=1,\qquad w_P(S)\le v_i(S).
\]
We delete goods of coefficient zero from $P$. 
Removing these goods preserves the pairwise disjointness of each agent's MMS bundles.
Their additive coefficients stay fixed as
we allocate goods; all guarantees refer to this original choice of
MMS partitions and lower bounds.

Whenever a pair of available goods gives an agent value at least
$\alpha$, we allocate the pair to her and remove the agent and the two
goods. Let $r$ denote the number of agents left when no such pair
exists. If $r=0$, we are done. Otherwise, we have allocated $n-r$
pairs and must account for their effect on each remaining agent's
bundles. Let $M'\subseteq M$ denote the goods left by these allocations.

For each remaining agent, we assign a nonnegative number
$\gamma_P$, called its {\em credit}, to each of her MMS bundles $P$.
For an unchanged bundle, let its credit be $\gamma_P=1$. For a
bundle losing one good, we keep its remaining goods and coefficients
and let $\gamma_P=1/2$. 
We assign credit zero to bundles losing at least two goods during the pair allocations and discard them. We call the other bundles \emph{surviving} bundles.
A bundle losing $\ell$ goods loses at most $\ell/2$ of its
initial credit. Since the original MMS bundles are disjoint, each
allocated good reduces their total credit by at most $1/2$.
As the initial total credit is \(n\), after allocating \(2(n-r)\) goods, every remaining agent satisfies
\begin{equation}\label{xos:eq:preprocessing-credit}
 \sum_P\gamma_P\ge r.
\end{equation}
For each remaining agent, her surviving bundles remain pairwise disjoint. Their additive values are
one when $\gamma_P=1$ and greater than $1-\alpha=11/16$ when
$\gamma_P=1/2$, since a removed coefficient is less than $\alpha$.
From now on, we continue to denote each surviving bundle's
remaining goods by $P$, with their fixed coefficients unchanged.

The pair allocations give us one further useful property. For every remaining agent $i$, every surviving bundle \(P\) of agent \(i\), and every two distinct available goods \(g,h\in M'\),
\begin{equation}\label{xos:eq:pair-bound}
 v_i(\{g,h\})<\alpha,\qquad w_P(g)+w_P(h)<\alpha.
\end{equation}
In the rest of the proof, we use these coefficient bounds, the
credits, and the disjointness of the surviving bundles.

\subsection{Coordinating the reserved goods}
\label{xos:subsec:options}
\label{xos:subsec:potential}

We prove a general reservation theorem on a finite set $M$,
which we apply to the remaining goods $M'$ in
Section~\ref{xos:subsec:menus}.
We first prove the rounding guarantee used to determine the required auxiliary value.
For a nonnegative additive valuation $\widetilde v$ on a finite set $M$, let
$\widetilde v(g)$ denote its coefficient on good $g$, and let
$\lVert \widetilde v\rVert_\infty=\max_{g\in M}\widetilde v(g)$, with value zero when $M$
is empty. A fractional allocation consists of numbers $f_{ig}\ge0$
such that $\sum_i f_{ig}\le1$ for every good $g$.

\begin{lemma}\label{xos:lem:additive-rounding}
Let $\widetilde v_i$ be nonnegative additive valuations on a finite set $M$, and
let $f$ be a fractional allocation. There are pairwise disjoint bundles
$B_i\subseteq M$ such that every agent $i$ satisfies
\[
 \widetilde v_i(B_i)\ge\sum_{g\in M}\widetilde v_i(g)f_{ig}-\lVert \widetilde v_i\rVert_\infty.
\]
\end{lemma}
We prove Lemma~\ref{xos:lem:additive-rounding} in
Appendix~\ref{xos:app:rounding}, following Shmoys and Tardos~\cite{ST93}.

Suppose an agent with valuation $v$ needs value $\beta>0$. We choose a
reserved set $H$, nonnegative additive coefficients $\widetilde v$ supported
outside $H$, and a threshold $\tau\in\R$ such that, for every
$B\subseteq M\setminus H$,
\begin{equation}\label{xos:eq:target}
 \widetilde v(B)>\tau-\lVert \widetilde v\rVert_\infty
 \quad\Longrightarrow\quad v(H\cup B)\ge \beta.
\end{equation}
We call such a triple $(H,\widetilde v,\tau)$ a {\em choice} and select all three components together.
If the agent receives \(H\) and obtains fractional auxiliary value greater than \(\tau\) on the unreserved goods, Lemma~\ref{xos:lem:additive-rounding} gives an integral bundle \(B\subseteq M\setminus H\) satisfying \(\widetilde v(B)>\tau-\|\widetilde v\|_\infty\).
By (\ref{xos:eq:target}), her final bundle \(H\cup B\) has value at least \(\beta\). 
We must choose a triple \((H_i,\widetilde v_i,\tau_i)\) for each agent so that the reserved sets are pairwise disjoint and every agent's fractional auxiliary value exceeds her threshold. 
The same reserved set may occur in different choices with different auxiliary valuations and thresholds.


To coordinate these choices, we use exponential random variables.
They let us interpret an increase in an expected maximum as
fractional auxiliary value. Let $Z$ be the mean-one exponential random variable
(i.e., $\Prb(Z>t)=e^{-t}$ for $t\ge0$ and $\E Z=1$).
Let $Z_1,\ldots,Z_k$ be independent copies of $Z$, also independent of $Z$.
For nonnegative
coefficients $b_1,\ldots,b_k$, let $V=\max_j b_jZ_j$, with an empty
maximum equal to zero, and, for $x\ge0$, let
\[
 \mathcal R_V(x)=\E(xZ-V)_+.
\]
Since \(\max\{V,xZ\}-V=(xZ-V)_+\), the quantity \(\mathcal R_V(x)\) is the increase in the expected maximum when \(xZ\) is added.
Conditioning on \(V\) and using the exponential distribution, we obtain, for \(x>0\),
\begin{equation}\label{xos:eq:memoryless}
 \mathcal R_V(x)=x\Prb(xZ>V).
\end{equation}
Thus, assigning a fractional share \(\mathbb P(xZ>V)\) of a good with auxiliary coefficient \(x\) gives auxiliary value \(\mathcal R_V(x)\), equal to the increase in the expected maximum.

Let $r\ge1$ agents have monotone valuations $v_i$ on a finite set $M$,
with targets $\beta_i>0$.
For each agent $i$, we will specify a finite collection of choices $(H_i,\widetilde v_i,\tau_i)$ satisfying \eqref{xos:eq:target} for her valuation $v_i$ and target $\beta_i$.
Here $H_i\subseteq M$, the auxiliary
coefficients are nonnegative and supported outside $H_i$, and
$\tau_i\in\R$. We also allow the empty choice $(\varnothing,0,0)$,
so that we can omit an agent temporarily when comparing her choices.
Choices with $H_i=\varnothing$ may still have nonzero auxiliary
coefficients.

Use one independent mean-one exponential $Z_j$ for each agent $j$.
We use the distributions of these variables to define
expectations and fractional shares; no sampling is needed.
For fixed choices of the agents other than $i$, let
\[
 b_g=\sum_{j\ne i}\widetilde v_j(g),\qquad
 V_g=\max_{j\ne i}\widetilde v_j(g)Z_j.
\]
The quantity $b_g$ is their total auxiliary coefficient on $g$ and
bounds $\E V_g$. All coefficients of a
choice are fixed in advance.
If another agent reserves a good, a coefficient on that good is
neither rescaled nor redistributed; it is simply unavailable in the
fractional allocation.

The following theorem gives a sufficient condition under which these choices can be coordinated to meet every agent's target.

\begin{theorem}[Reservation theorem]\label{xos:thm:reservation}
For every agent $i$ and every choice of the other agents with
pairwise disjoint reserved sets, including empty choices, let
$H=\bigcup_{j\ne i}H_j$. Suppose agent $i$ always has a choice with
$H_i\cap H=\varnothing$ for which
\begin{equation}\label{xos:eq:surplus}
 \sum_{g\in M\setminus(H\cup H_i)}\mathcal R_{V_g}(\widetilde v_i(g))
 -\tau_i-\sum_{g\in H_i}b_g>0.
\end{equation}
Then there is an allocation giving every agent $i$ value at least
$\beta_i$.
\end{theorem}
\begin{proof}
For choices with disjoint reserved sets, let $\Phi$ denote the
potential
\begin{equation}\label{xos:eq:potential}
 \Phi=\sum_{g\in M\setminus\bigcup_jH_j}\E\max_j \widetilde v_j(g)Z_j
       -\sum_j\tau_j.
\end{equation}
We choose triples maximizing $\Phi$. Such choices exist because the collections are finite and the all-empty
choice is allowed. Every expectation is finite, since the maximum is
at most the sum.

Fix an agent $i$, let $H$ be the other agents' reserved goods, and
temporarily replace her choice by the empty one.
Inserting a choice supplied by the hypothesis changes
$\Phi$ by the left side of~\eqref{xos:eq:surplus} with $\E V_g$ in
place of $b_g$. Indeed, a good reserved by $i$ leaves the expected
maximum, costing $\E V_g$, while each other unreserved good gains
$\mathcal R_{V_g}(\widetilde v_i(g))$. 
Since \(\mathbb EV_g\le b_g\), this change is positive. 
By maximality, the current choice of agent \(i\) gives potential at least as large as this alternative choice. 
Therefore, her current choice also contributes positively relative to the empty choice.

To finish, we show that these choices give every agent enough
fractional auxiliary value for rounding. For agent $i$, let
\[
 U_i=\sum_{g\in M\setminus\bigcup_jH_j}\mathcal R_{V_g}(\widetilde v_i(g)).
\]
The contribution of the current choice is exactly
$U_i-\tau_i-\sum_{g\in H_i}\E V_g$. This remains true if some of
its auxiliary coefficients lie on other agents' reserved goods: those
goods are excluded before and after removing her choice. Hence
$U_i>\tau_i$, and her choice is not the additional empty one.

For every unreserved good \(g\) and agent \(i\) with \(\widetilde v_i(g)>0\), let \(f_{ig}\) be the probability that \(\widetilde v_i(g)Z_i\) is strictly larger than the corresponding products of all other agents. Set \(f_{ig}=0\) when \(\widetilde v_i(g)=0\). Since positive ties have probability zero, these probabilities form a fractional allocation on the unreserved goods. 
By~\eqref{xos:eq:memoryless}, the
auxiliary value of agent $i$ is $U_i$.
Lemma~\ref{xos:lem:additive-rounding} gives disjoint bundles $B_i$ of
unreserved goods with
\[
 \widetilde v_i(B_i)\ge U_i-\lVert \widetilde v_i\rVert_\infty
          >\tau_i-\lVert \widetilde v_i\rVert_\infty.
\]
Here $\lVert \widetilde v_i\rVert_\infty$ bounds the largest coefficient
available for rounding. We may remove from each rounded bundle
$B_i$ any good with zero auxiliary coefficient for agent $i$,
without changing the rounding inequality. Equation~\eqref{xos:eq:target} gives
$v_i(H_i\cup B_i)\ge \beta_i$ for every agent. The rounded bundles $B_i$
and the reserved sets $H_i$ are mutually disjoint. Any leftover goods
may be allocated by monotonicity.
\end{proof}

\paragraph{Applying the reservation theorem.}
In our application, we verify the hypothesis using disjoint bundles
and their credits. Let $a>0$ and suppose every choice satisfies
\begin{equation}\label{xos:eq:option-budget}
 \widetilde v_i(M)+a|H_i|\le1.
\end{equation}
As proved in Section~\ref{xos:subsec:reduction}, the pair
allocations leave each remaining agent with a finite family of
pairwise disjoint surviving MMS bundles $P\subseteq M$ whose
nonnegative credits $\gamma_P$ satisfy
\begin{equation*}
 \sum_P\gamma_P\ge r.
\end{equation*}
Fix any agent $i$ and choices of the others as in the theorem, and
let $H=\bigcup_{j\ne i}H_j$. For each of her bundles $P$, let
$\ell=|P\cap H|$ and $L=\sum_{g\in P\setminus H}b_g$.
It suffices to show that, whenever
\begin{equation}\label{xos:eq:local-condition}
 L<\gamma_P-a\ell,
\end{equation}
she has a choice satisfying~\eqref{xos:eq:surplus} whose reserved
set and auxiliary support lie in $P\setminus H$.
Indeed, her disjoint bundles count each reserved good at most once
and each auxiliary coefficient of another agent at most once.
Consequently,
\begin{align}\label{xos:eq:credit-balance}
 \sum_P\left(\gamma_P-a|P\cap H|
       -\sum_{g\in P\setminus H}b_g\right)
 &\ge r-a|H|-\sum_{j\ne i}\widetilde v_j(M)\nonumber\\
 &\ge1.
\end{align}
The last inequality sums the $r-1$ bounds
in~\eqref{xos:eq:option-budget}, since their reserved sets are
disjoint. Thus some bundle satisfies~\eqref{xos:eq:local-condition}.
It remains to construct the choices and verify this implication for \(a=1/3\).

\subsection{Choosing the auxiliary valuations}
\label{xos:subsec:coefficients}

We first construct auxiliary valuations from fixed additive lower bounds and then estimate their fractional values using the other agents' coefficients.
Capping preserves the lower-bound property, but subsequent scaling may increase the coefficients, so the auxiliary valuation need not be a lower bound on the original valuation.
We choose the threshold of each choice so that~\eqref{xos:eq:target} accounts for the loss of the largest auxiliary coefficient in rounding.


\begin{lemma}
\label{xos:lem:capped-tail}
Let $\beta>0$, let $w\le v$ be a nonnegative additive lower bound on $M$, and let
$H\subseteq M$ satisfy $w(H)<\beta$. Let $z\le w$ be nonnegative additive
coefficients supported outside $H$, with $z(M)>0$. For every
$\lambda>0$, let
\[
 \widetilde v=\frac{\lambda z}{z(M)}
\]
and
\[
\tau=\lambda\frac{\beta-w(H)+\lVert z\rVert_\infty}{z(M)}.
\]
Then $\widetilde v(M)=\lambda$ and $(H,\widetilde v,\tau)$ satisfies~\eqref{xos:eq:target}.
\end{lemma}
\begin{proof}
By construction, \(\widetilde v(M)=\lambda\).
Fix \(B\subseteq M\setminus H\) satisfying \(\widetilde v(B)>\tau-\|\widetilde v\|_\infty\). By the definitions of \(\widetilde v\) and \(\tau\), and since \(\|\widetilde v\|_\infty=\lambda\|z\|_\infty/z(M)\), 
we have $z(B)>\beta-w(H)$.
Thus,
\[
v(H\cup B)\ge w(H)+w(B)\ge w(H)+z(B)>\beta,
\]
where the first inequality follows from \(w\le v\), the additivity of \(w\), and \(H\cap B=\varnothing\).
This proves~\eqref{xos:eq:target}.
\end{proof}

For choices without reserved goods, we next bound the required threshold using the additive value remaining after removing the largest coefficient. 
We cap the coefficients to control the rounding loss while ensuring that sufficiently large auxiliary value implies original additive value at least \(\beta\).


\begin{lemma}[Cap and restore]\label{xos:lem:cap}
Let $w$ be nonnegative additive coefficients on a nonempty finite set
$P$, and let $A=w(P)-\max_{g\in P}w(g)$. If $0<\beta<A$, there are
nonnegative auxiliary coefficients $\widetilde v$ on $P$ with $\widetilde v(P)=1$ such that
\begin{equation}\label{xos:eq:cap}
 \lVert \widetilde v\rVert_\infty+
 \max_{B\subseteq P:\,w(B)<\beta}\widetilde v(B)\le\frac A{2A-\beta}.
\end{equation}
In particular, if $w\le v$, the choice with no reserved goods and
threshold $A/(2A-\beta)$ satisfies~\eqref{xos:eq:target}.
\end{lemma}
\begin{proof}
Omit zero coordinates, let $N$ be the number of positive coordinates,
and let
\[
 \kappa=1+\frac A{A-\beta}>2.
\]
The bound we need is $(\kappa-1)/\kappa$.
If $N\le\kappa$, let every auxiliary coefficient be $1/N$. Every set
of $N-1$ goods has original value at least $A>\beta$, so a set of original
value less than $\beta$ contains at most $N-2$ goods. The left side of
\eqref{xos:eq:cap} is at most
$(N-1)/N\le(\kappa-1)/\kappa=A/(2A-\beta)$.

Suppose $N>\kappa$, and let $g_0$ have largest coefficient. We choose
a cap so that, when this good is capped, the total capped value is
$\kappa$ times the cap. Let $z>0$ satisfy
\begin{gather}
 \sum_{g\ne g_0}\min\{w(g),z\}=(\kappa-1)z,\label{xos:eq:waterfill}
\end{gather}
and
\[
 z\le\frac A{\kappa-1}=A-\beta.
 \]
Such a cap exists by continuity: the difference between the two sides
is positive for sufficiently small positive $z$, because $N>\kappa$,
and is nonpositive at $z=A/(\kappa-1)$.
If $w(g_0)<z$, let $\widetilde v=w/w(P)$. Then
\[
 \lVert \widetilde v\rVert_\infty+\max_{w(B)<\beta}\widetilde v(B)
 \le\frac{w(g_0)+\beta}{A+w(g_0)}\le\frac A{2A-\beta},
\]
where the last inequality follows from $w(g_0)< z\le A-\beta$.

It remains to consider $w(g_0)\ge z$. Cap every coefficient at $z$
and normalize by their total $\kappa z$. The largest auxiliary
coefficient is $1/\kappa$. To prove~\eqref{xos:eq:cap}, it suffices
to show that any set with capped value greater than $(\kappa-2)z$
has original value at least $\beta$. Let $k$ be the number of
capped goods, those with original coefficient at least $z$, and let
$T$ be the total original value of the other goods.
Equation~\eqref{xos:eq:waterfill} gives $T=(\kappa-k)z$. 
Since $g_0$
is capped, the $k-1$ smallest original coefficients among the capped
goods sum to $A-T$.
A set with capped value greater than $(\kappa-2)z$ contains at least
$k-1$ capped goods: at most $k-2$ such goods together with all goods
whose coefficients are below $z$ have capped value at most
$(k-2)z+T=(\kappa-2)z$. For $k=1$, the assertion about the number
of members is automatic. Restoring the original coefficients of
these $k-1$ goods adds at least $A-T-(k-1)z$. The original value
of the set is therefore greater than
\[
 (\kappa-2)z+A-T-(k-1)z=A-z\ge \beta.
\]
Thus a set of original value less than $\beta$ has auxiliary value at
most $(\kappa-2)/\kappa$. The largest auxiliary coefficient is
$1/\kappa$, proving~\eqref{xos:eq:cap}. Extend $\widetilde v$ by zero on omitted
coordinates.

Finally, if \(w\le v\), inequality (\ref{xos:eq:cap}) implies that every set \(B\) satisfying \(\widetilde v(B)>A/(2A-\beta)-\|\widetilde v\|_\infty\) has \(w(B)\ge\beta\), and hence \(v(B)\ge\beta\). 
Thus, the choice with \(H=\varnothing\) and \(\tau=A/(2A-\beta)\) satisfies (\ref{xos:eq:target}).
\end{proof}

We now bound the auxiliary value obtained against competing choices.
The first two estimates below will let us combine bounds on the
potential increases from choices that reserve different goods.
The last estimate uses only the sum of
the other agents' coefficients when the auxiliary coefficients sum to one.

\begin{lemma}\label{xos:lem:response}
Let $(b_j)_j$ be a finite family of nonnegative coefficients and let
$V=\max_jb_jZ_j$, with independent exponential variables as in
\eqref{xos:eq:memoryless}. For every $x\ge0$,
\begin{equation}
    \begin{aligned}
 &\mathcal R_V(x)\ge\frac34x-\frac14\sum_jb_j,\\
 &\mathcal R_V(x)\ge\frac{73}{80}x-\frac12\sum_jb_j.\label{xos:eq:two-affines}
\end{aligned}
\end{equation}

For the quantities $b_g,V_g$ defined above, let $\widetilde v$ be a nonnegative
auxiliary valuation with $\widetilde v(M)=1$. If
$L=\sum_{g\in\supp(\widetilde v)}b_g\le1$, then
\begin{equation}\label{xos:eq:nonlinear}
 \sum_{g\in M}\mathcal R_{V_g}(\widetilde v(g))\ge\frac1{1+L}.
\end{equation}
\end{lemma}

We prove these estimates in Appendix~\ref{xos:app:exponential}.
Combining~\eqref{xos:eq:nonlinear} with Lemma~\ref{xos:lem:cap}
gives a condition that will handle bundles whose value remains
large after removing their largest coefficient.

\begin{corollary}\label{xos:cor:tail}
Let $w\le v$ be nonnegative additive coefficients supported on
a nonempty set $P\subseteq M$, and let $A=w(P)-\max_{g\in P}w(g)$.
If $0\le L<1$ and $0<\beta<(1-L)A$, the choice from
Lemma~\ref{xos:lem:cap} has total expected increase
$\sum_{g\in P}\mathcal R_{V_g}(\widetilde v(g))$ strictly greater
than its threshold whenever the other agents' coefficients
sum to at most $L$ on its support.
\end{corollary}
\begin{proof}
The assumption gives $A>\beta$ and $A/(2A-\beta)<1/(1+L)$.
Extend the auxiliary valuation by zero outside $P$ and apply
\eqref{xos:eq:nonlinear} to the choice from
Lemma~\ref{xos:lem:cap}, whose coefficients sum to one.
\end{proof}
\subsection{The choices available to each agent}
\label{xos:subsec:menus}

We apply the reservation theorem to the $r$ remaining agents and the
goods $M'$, with each $v_i$ restricted to $M'$. The target is
$\alpha=5/16$, measured against the original MMS benchmark of one.
The charge for reserving one good will be $1/3$.
This is the largest charge compatible with the unit budget
in~\eqref{xos:eq:option-budget} when we reserve three goods and use
zero auxiliary coefficients.
Thus a choice that reserves one good may use auxiliary coefficients
of total $2/3$, while a choice with no reserved goods may use total
one.

To allow an agent to avoid goods reserved by the others, we
apply the following constructions to every subset of each surviving
MMS bundle, using its fixed coefficients restricted to that subset.
Each resulting triple uses reserved goods and auxiliary coefficients
from the same subset, and hence from one original MMS bundle.
All coefficients and caps are fixed when the choices are constructed.
We also fix an ordering of goods to resolve ties.

\paragraph{Reserving no good.}
For a restriction with at least two positive coefficients, let $R$
be their total and let $h\ge s$ be the two largest coefficients.
We cap the largest coefficient at $s$ and normalize the resulting
coefficients to sum to one. Let $\tau_0$ denote the threshold
supplied by Lemma~\ref{xos:lem:capped-tail}, namely
\begin{equation}\label{xos:eq:flat}
 \tau_0=\frac{\alpha+s}{R-h+s}.
\end{equation}
If $R-h>\alpha$, we also include the choice given by
Lemma~\ref{xos:lem:cap}. We fix its cap for each restriction before
considering the other agents' choices.

\paragraph{Reserving one good.}
Suppose the restriction has at least three positive coefficients,
with largest values $h\ge s\ge t$, and $h+s<\alpha$.
We consider two possible choices, each with auxiliary coefficient sum \(2/3\).
For the first, reserve the good of coefficient $h$ and cap the
coefficient $s$ at $z=\min\{s,\alpha-h-s\}$. 
We include this choice when \(t\le z\).
For the second,
reserve the good of coefficient $s$ and omit the good of
coefficient $h$ from the auxiliary valuation. Normalize the
remaining coefficients in each choice to sum to $2/3$. Let
$\tau_h,\tau_s$ denote their thresholds. By
Lemma~\ref{xos:lem:capped-tail}, we may take
\begin{equation*}
 \tau_h=\frac23\frac{\alpha-h+z}{R-h-s+z}, \quad \tau_s=\frac23\frac{\alpha-s+t}{R-h-s}.
\end{equation*}

\paragraph{Reserving three goods.}
If three coefficients sum to at least $\alpha$, reserve their goods and
take $\widetilde v=0$ and $\tau=-1$. The reserved set alone meets the target.
The choice contributes one through the threshold term in the
potential, before accounting for the other agents' coefficients on
its reserved goods.

These choices are finite in number, since the bundle families and
their subsets are finite and every cap is fixed in advance. They
all satisfy the budget in~\eqref{xos:eq:option-budget}: the totals
for zero, one, and three reserved goods are $1$, $2/3+1/3$, and
$3/3$, respectively.

Algorithm~\ref{xos:alg:reservation} summarizes the construction in the
proof of Theorem~\ref{xos:thm:reservation}, using these choice families
on the remaining goods $M'$.

\begin{algorithm}
\caption{Reservation and rounding}\label{xos:alg:reservation}
\begin{algorithmic}[1]
\Require Remaining goods $M'$ and finite families of choices
$(H_i,\widetilde v_i,\tau_i)$ constructed above for $r\ge1$ remaining agents,
indexed by $1,\ldots,r$.
\State Add the empty choice $(\varnothing,0,0)$ to every family.
\State Let $Z_1,\ldots,Z_r$ be independent mean-one exponential random variables.
\State Enumerate all combinations of one choice per agent with pairwise
disjoint reserved sets and choose one
maximizing the potential in~\eqref{xos:eq:potential} on $M'$.
\For{each unreserved good $g\in M'\setminus\bigcup_jH_j$}
 \State For every agent $i$, let
 \[
 f_{ig}=\Prb\!\left[\widetilde v_i(g)Z_i>
          \max_{j\ne i}\widetilde v_j(g)Z_j\right],
 \]
 with an empty maximum equal to zero.
\EndFor
\State Enumerate pairwise disjoint bundles $B_i$ of unreserved
goods with positive auxiliary coefficients for their recipients until,
for every agent $i$,
\[
 \widetilde v_i(B_i)\ge
 \sum_{g\in M'\setminus\bigcup_jH_j}\widetilde v_i(g)f_{ig}
 -\lVert\widetilde v_i\rVert_\infty.
\]
\State \Return the bundles $H_i\cup B_i$, distributing unassigned goods arbitrarily.
\end{algorithmic}
\end{algorithm}
Before distributing leftover goods, each remaining agent's reserved and
rounded goods lie in the one original MMS bundle from which her
selected triple was constructed.

\subsection{Completing the allocation}
\label{xos:subsec:local}

The credit argument following Theorem~\ref{xos:thm:reservation} guarantees that, for every remaining agent and every choice of the other agents with pairwise disjoint reserved sets, some surviving bundle satisfies~\eqref{xos:eq:local-condition} with \(a=1/3\).
We now show that every such bundle provides a choice satisfying (\ref{xos:eq:surplus}).

Fix a remaining agent \(i\), a surviving bundle \(P\) of agent \(i\), and choices of the other agents whose reserved sets are pairwise disjoint. 
Let \(H\) be the goods reserved by the other agents and let \(b_g\) be the sum of their auxiliary coefficients on good \(g\). 
Let \(\ell=|P\cap H|\) and \(L=\sum_{g\in P\setminus H}b_g\). 
Here \(\ell\) counts the unavailable goods of \(P\), and \(L\) measures the other agents' auxiliary coefficients on its remaining goods.


\begin{lemma}\label{xos:lem:local}
For every surviving bundle $P$, if $L<\gamma_P-\ell/3$, then its
owner has an available choice inside $P\setminus H$ satisfying
\eqref{xos:eq:surplus}.
\end{lemma}
\begin{proof}
Since $L\ge0$, we have $\ell\in\{0,1,2\}$ for credit one and
$\ell\in\{0,1\}$ for credit one half. Removing these $\ell$ goods leaves
additive value greater than $\alpha$: the bundle started with value
one or more than $1-\alpha$, and every removed coefficient is less
than $\alpha$. By the pair bound~\eqref{xos:eq:pair-bound}, at least three
positive coefficients remain. Let $R$ be their total and let
$h\ge s\ge t$ be the three largest.
By (\ref{xos:eq:pair-bound}), we have \(h+s<\alpha\), and hence \(s<\alpha/2\).

We first show that individual choices settle every situation except
a bundle of credit one half with no goods reserved by the
other agents. We then combine three gain bounds to handle that
remaining situation.

\paragraph{When one choice suffices.}
Suppose, towards a contradiction, that no available choice satisfies
\eqref{xos:eq:surplus}. The choice with no reserved goods has
threshold $\tau_0$ from~\eqref{xos:eq:flat} and obtains auxiliary
fractional value at least $1/(1+L)$ by~\eqref{xos:eq:nonlinear}.

For a bundle of credit one, the \(\ell\) goods reserved by the other agents have total coefficient \(1-R\) under \(w_P\).
We claim that $1-R+h+L(\alpha+s)<2\alpha$.
For $\ell=0$, this follows from $h+s<\alpha$ and $L<1$. For $\ell\ge1$,
the coefficients of the $\ell$ reserved goods and $h$ have pairwise
sums below $\alpha$. Their total is less than $\alpha(\ell+1)/2$. Since
$s<\alpha/2$ and $L<1-\ell/3$, it follows that
\[
 1-R+h+L(\alpha+s)
 <\frac\alpha2(\ell+1)+\frac{3\alpha}{2}
       \left(1-\frac\ell3\right)=2\alpha.
\]
Consequently,
\[
 R-h+s-(1+L)(\alpha+s)
 =R-h-\alpha-L(\alpha+s)>1-3\alpha=\frac1{16}.
\]
Thus $\tau_0<1/(1+L)$, and this choice satisfies
\eqref{xos:eq:surplus}. The bundle must therefore have credit one
half.

If $\ell=1$, one good was removed during the pair allocations and
one is now reserved by an opponent. The former coefficient is less
than $\alpha$. By~\eqref{xos:eq:pair-bound}, the latter coefficient
together with $h$ is less than $\alpha$. Hence $L<\frac16$ and
\[
 R-h>1-2\alpha=\frac38.
\]
These estimates explain our target: the sufficient inequality
$\alpha\le\frac56(1-2\alpha)$ is equivalent to $\alpha\le5/16$.
At $\alpha=5/16$, the strict bounds above give
$(1-L)(R-h)>\frac56(1-2\alpha)=\alpha$.
Corollary~\ref{xos:cor:tail} therefore supplies
a choice satisfying~\eqref{xos:eq:surplus}. We must have $\ell=0$.

We are left with $R>11/16$ and $L<1/2$. If $h+s+t\ge\alpha$, reserving
these three goods gives a left side in~\eqref{xos:eq:surplus} of
at least $1-L>0$. Also, failure of the choice with no reserved
goods implies $\tau_0\ge1/(1+L)>2/3$. Thus it suffices to handle
$\gamma_P=\frac12$, $\ell=0$, $h+s+t<\alpha$ and $\tau_0>\frac23$.

\paragraph{Combining three gain bounds.}
The sum of the two largest coefficients falls short of the target
by $\alpha-h-s$; let $\delta=\alpha-h-s$ denote this remaining value.
We have $t<\delta$ and
$R-h-s>3/8+\delta$.
The inequality $\tau_0>2/3$ gives
$2h+s>2R-3\alpha>7/16$. Since
$2h+s=2\alpha-2\delta-s$, we obtain
$0<\delta<3/32$ and $0\le s<3/16-2\delta$.

We compare the three choices from Section~\ref{xos:subsec:menus}
that reserve no good, the good of coefficient $h$, or the good
of coefficient $s$.
Their thresholds are $\tau_0,\tau_h,\tau_s$. Since
$t<\delta$ and $t\le s$, we have $t\leq z=\min\{s,\delta\}$, so both choices with one reserved good are available.
Their thresholds satisfy
\begin{align}
 \tau_h\le\frac23\frac{2\delta+s}{3/8+2\delta}, \quad \tau_s\le\frac23\frac{\alpha+\delta-s}{3/8+\delta}.\label{xos:eq:headed-bounds}
\end{align}
For the first bound, use $R-h-s>3/8+\delta$ and increase $z$ to
$\delta$. The ratio $(\delta+s+z)/(3/8+\delta+z)$ increases with
$z$ because $s<3/16<3/8$. The second bound follows from
$t<\delta$ and $R-h-s>3/8+\delta$.

Let $\Delta_0,\Delta_h,\Delta_s$ denote the left side of
\eqref{xos:eq:surplus} for these three choices. It suffices to show
that a convex combination of these numbers is positive. Let
$b_h,b_s$ denote the other agents' total auxiliary coefficients on
the goods with coefficients $h,s$. We choose the coefficients
of the convex combination so that the terms in $b_h$ and $b_s$
cancel. For the choice with no reserved
goods, both of its largest auxiliary coefficients equal
$s/(R-h+s)$. Apply the first bound in~\eqref{xos:eq:two-affines}
to these two goods and the second bound to all other goods.
For the two choices with one reserved good, apply the second bound
on their entire auxiliary supports. Subtracting the other agents'
coefficients on each reserved good, as in~\eqref{xos:eq:surplus},
we obtain
\begin{align*}
 \Delta_0&\ge\frac{73}{80}-\tau_0
       -\frac{13s}{40(R-h+s)}-\frac12L
       +\frac14(b_h+b_s),\\
 \Delta_h&\ge\frac{73}{120}-\tau_h-\frac12L-\frac12b_h,\\
 \Delta_s&\ge\frac{73}{120}-\tau_s-\frac12L
       +\frac12b_h-\frac12b_s.
\end{align*}
Reserving a good costs its full competing coefficient,
while omitting it from the auxiliary support saves half that
coefficient in the bound on fractional loss. The last choice omits
both largest goods and reserves the second, so the terms involving
$b_h,b_s$ are $(b_h+b_s)/2-b_s=b_h/2-b_s/2$, as in the last line.

Cancelling $b_s$ requires twice as much weight on $\Delta_0$
as on $\Delta_s$. Cancelling $b_h$ then requires equal weights on
$\Delta_0$ and $\Delta_h$. The ratio is therefore $2:2:1$, giving
\begin{equation}\label{xos:eq:mixture}
 \frac25\Delta_0+\frac25\Delta_h+\frac15\Delta_s.
\end{equation}
Since $L<1/2$, this combination is strictly greater than
\[
 \frac{12}{25}-\frac25\tau_0-\frac25\tau_h
 -\frac15\tau_s-\frac{13s}{100(R-h+s)}.
\]
Substituting~\eqref{xos:eq:headed-bounds} and
$R-h+s>3/8+\delta+2s$ shows that it is positive if
\begin{equation}\label{xos:eq:threshold-balance}
 \frac{1/8+(53/100)s}{3/8+\delta+2s}
 +\frac{4(2\delta+s)}{15(3/8+2\delta)}
 +\frac{2(\alpha+\delta-s)}{15(3/8+\delta)}<\frac{12}{25}.
\end{equation}

For fixed \(\delta\), the left-hand side of (\ref{xos:eq:threshold-balance}) is convex in \(s\), so it suffices to check the endpoints \(s=0\) and \(s=3/16-2\delta\).
In Appendix~\ref{xos:app:scalar}, we verify~\eqref{xos:eq:threshold-balance} for \(0\le\delta\le3/32\) and \(0\le s\le3/16-2\delta\).
Thus the convex combination
in~\eqref{xos:eq:mixture} is positive. At least one of the three
choices satisfies~\eqref{xos:eq:surplus}, completing the proof.
\end{proof}

\begin{proof}[Proof of Theorem~\ref{thm:xos-main}]
We apply the singleton reduction and normalize each agent's
MMS to one, as in
Section~\ref{xos:subsec:reduction}. Each pair allocated there
gives its recipient value at least $\alpha$. If no agents remain, we are done.
For the agents who remain,
\eqref{xos:eq:preprocessing-credit} gives enough total credit.
The choices constructed in Section~\ref{xos:subsec:menus} satisfy
the unit budget with charge $1/3$ per reserved good, and
Lemma~\ref{xos:lem:local} supplies a choice whenever
\eqref{xos:eq:local-condition} holds. By the credit argument
in~\eqref{xos:eq:credit-balance}, such a bundle exists against every
choice of the other agents with disjoint reserved sets.
Theorem~\ref{xos:thm:reservation}
therefore guarantees that Algorithm~\ref{xos:alg:reservation} gives every
remaining agent value at least $\alpha$. Combining its output with the allocated
pairs gives an $\alpha$-MMS allocation
on the working instance. The singleton reduction then gives the
required $5/16$-MMS allocation in the original instance.
\end{proof}

\section{Conclusion}\label{sec:conclusion}

We improve the MMS guarantee for $n$ agents with general subadditive valuations to $\Omega(1/\log^* n)$ through a new amplification theorem for bidding under fixed budgets. 
Since the anyprice share (APS) is at most five times the MMS for subadditive valuations~\cite{FG25}, our guarantees also hold for APS with a loss of at most a factor of five.
We also prove constant guarantees for $q$-partitioning valuations whenever $q/n$ is bounded below by a positive constant. 
These guarantees use supporting prices on groups and extend beyond XOS whenever \(q<m\).
At the XOS endpoint, item-level additive lower bounds allow us to improve the universal guarantee further to $5/16$. Our reservation theorem coordinates integral assignments of selected goods with a fractional allocation of the remaining goods, while accounting for the loss in rounding.

The main open question is whether every instance with subadditive valuations admits an allocation giving each agent a universal constant fraction of her MMS. For $q$-partitioning valuations, we ask whether a constant guarantee can be obtained for some $q=o(n)$, with $q=2$ corresponding to the general subadditive question. Another question is whether group supporting prices can replace the item-level additive lower bounds in our reservation argument, giving stronger guarantees for intermediate classes. For XOS valuations, the gap between our $5/16$ guarantee and the known $1/2$ upper bound remains open~\cite{GHS+22}.
We also ask whether our reservation theorem can improve APS guarantees for XOS valuations without the loss from converting MMS guarantees.


\appendix
\section{An initialization using only bidding}\label{app:bidding}
The bound of \(1+\log_2^*n\) layers in Proposition~\ref{prop:layers} uses Corollary~\ref{cor:partial}, based on the partial-allocation theorem of Seddighin and Seddighin, to construct the first three layers.
This appendix gives an alternative initialization using only Theorem~\ref{thm:amplification}.
It shows that the iterated-logarithmic layer bound can be obtained using bidding amplification alone, at the cost of a larger additive constant. 
The exponential recurrence and the final rounding argument remain unchanged.


The amplified phase in Proposition~\ref{prop:layers} begins once $n/u\ge42$, where $u$ is the number of unserved agents. Initially $u=n$, so this ratio is one. 
We first apply one-trial bidding to groups of unserved agents until at most \(n/4\) remain.
We then apply one-trial bidding repeatedly to all unserved agents. After a constant number of layers, fewer than \(n/42\) agents remain, allowing us to begin the amplified phase.


\begin{proposition}\label{prop:bidding-only}
Suppose ${n}\ge2$ agents have normalized, monotone, subadditive valuations and each has ${n}$ disjoint sets of value at least one. 
Then there exists a multi-allocation giving every agent value at least \(1/2\), with multiplicity at most \(19+\log_2^*n\).
Every partial-allocation layer is obtained from \cref{thm:amplification} or by serving one agent directly.
\end{proposition}

\paragraph{The alternative layer procedure.}
Every layer is constructed on the full ground set. After recording a successful agent's bundle, remove the agent from later layers but leave the goods available. In each use of \cref{thm:amplification}, choose a realization satisfying its bound on the number of failures.
\begin{enumerate}
\item If ${n}\le4$, serve the agents one at a time, giving each any one of her sets of value at least one in a separate layer, and stop. For ${n}>4$, let $h=\lfloor {n}/4\rfloor+1$. While more than ${n}/4$ agents remain unserved, choose any $h$ of them and apply one-trial bidding to this group.
\item Once at most ${n}/4$ agents remain, use at most nine more layers. In each, apply one-trial bidding to all unserved agents if at least two remain; serve a lone agent directly; stop immediately if everyone is served.
\item If anyone remains after Step~2, use the amplified loop from \cref{prop:layers}: for $u\ge2$ unserved agents choose $\ell=\lfloor {n}/[4(u-1)]\rfloor$ trials, and serve a lone agent directly. Stop when no unserved agents remain.
\end{enumerate}
Each agent's output bundle is the one recorded in her successful layer. This output is a multi-allocation; a good may have been used in different layers. The proof bounds the number of layers, which also bounds the number of recorded bundles containing any one good.

\begin{proof}
If ${n}\le4$, the individual allocations in Step~1 use at most four layers, fewer than $19+\log_2^*{n}$. Assume from now on that ${n}>4$.

\paragraph{At most twelve layers to reach ${n}/4$ unserved agents.}
Initially, \(u=n\), so the condition \(4(u-1)\le n\) required by Theorem~\ref{thm:amplification} for one-trial bidding does not hold. 
We instead choose \(h=\lfloor n/4\rfloor+1\) agents.
Each of these agents still has $n$ disjoint bundles of value at least one because earlier layers do not remove goods. The inequality $4(h-1)\le {n}$ verifies the sufficient condition in \cref{thm:amplification} with $u=h$ and $\ell=1$. Consequently, one layer leaves at most $\lfloor3h/4\rfloor$ members of the selected group unserved. The number served in this layer is at least $h-\lfloor3h/4\rfloor=\lceil h/4\rceil$.

The choice of $h$ makes this progress a fixed fraction of the initial count ${n}$: since $h>{n}/4$, we have $\lceil h/4\rceil>{n}/16$. Agents outside the group receive nothing in the current layer and remain eligible for later layers. There are enough agents to choose the group whenever the loop condition $u>{n}/4$ holds. Indeed, an integer $u$ with this property satisfies $u\ge\lfloor {n}/4\rfloor+1=h$.

Every executed layer thus serves more than ${n}/16$ previously unserved agents. If twelve layers are needed, they serve more than $12{n}/16=3{n}/4$ agents in total, leaving fewer than ${n}/4$. The loop therefore finishes after at most twelve layers. If it finishes earlier, the condition $u\le {n}/4$ already holds.

\paragraph{At most nine further layers to reach ${n}/42$.}
Now all unserved agents can participate together in one-trial bidding. If their current number is $u\ge2$, then $u\le {n}/4$ implies $4(u-1)\le {n}$. The amplification theorem with $\ell=1$ gives a layer with at most $\lfloor3u/4\rfloor$ failures. Since only failing agents continue, their number is multiplied by at most $3/4$ in every such layer. If exactly one agent remains, a direct layer finishes the construction.

Suppose there is still an unserved agent after nine layers of this phase. None of those layers served a lone agent directly, so all nine used the factor-$3/4$ bound. Starting from at most ${n}/4$ agents, we obtain
\[
u\le\frac n4\left(\frac34\right)^9<\frac n{42}.
\]
Thus $u<n/42$ after these nine layers. Altogether, at most $12+9=21$ layers either serve everyone or give the ratio ${n}/u>42$ required for the amplified recurrence.

\paragraph{The remaining amplified layers.}
Let $k=\log_2^*{n}\ge3$. If anyone remains after the initialization, the ratio $z={n}/(12u)$ is greater than $7/2$. Step~3 uses the same number of trials as the amplified loop in \cref{prop:layers}, so it satisfies the same recurrence~\eqref{eq:residual-recurrence}. Starting from $z>3$, if agents remain after the first additional layer, this recurrence gives $z>3\cdot2^3=24>E_3$. 
Inductively, after $r\ge1$ additional layers, any remaining agents give $z\ge E_{r+2}$. 
Indeed, if \(z\ge E_{r+2}\) and another layer leaves agents unserved, (\ref{eq:residual-recurrence}) gives a new ratio of at least \(z2^z\ge2^{E_{r+2}}=E_{r+3}\).
If agents remained after $k-2$ additional layers, we would therefore have
\[
 z\ge E_k\ge n,
 \qquad\text{whereas}\qquad
 z=\frac{n}{12u}\le\frac n{12}<n,
\]
a contradiction. Thus at most $k-2$ further layers suffice, including a possible direct layer for a lone agent.

The total is at most $21+(k-2)=19+\log_2^*{n}$. As each layer is feasible, each good belongs to at most one recorded bundle per layer. This proves the stated multiplicity bound.
\end{proof}

The same rounding argument turns this alternative layer count into an MMS guarantee. The smallest power of two at least the stated layer bound supplies the dyadic multiplicity bound required by \cref{thm:rounding}.

\begin{corollary}
For $n\ge2$, let $\widetilde D(n)=2^{\lceil\log_2(19+\log_2^*n)\rceil}$. Every subadditive instance admits a $1/(4\widetilde D(n)-2)$-MMS allocation. The approximation factor satisfies
\[
 \frac{1}{4\widetilde D(n)-2}>
 \frac{1}{8(19+\log_2^*n)}.
\]
The multi-allocation is constructed using only Theorem~\ref{thm:amplification} and direct allocations to individual agents.
\end{corollary}
\begin{proof}
Let $D=\widetilde D(n)$ and $\alpha=1/(4D-2)$. Since this target is nonincreasing in $n$, the reduction in Section~\ref{sec:preliminaries} allows us to assume that every agent's MMS is one and every singleton has value below $\alpha$. Each agent then has $n$ disjoint MMS bundles of value at least one. By \cref{prop:bidding-only}, there is a multi-allocation giving each agent value at least $1/2$ with multiplicity at most $19+\log_2^*n\le D$.

Applying \cref{thm:rounding} gives pairwise disjoint bundles with value at least
\[
\frac{1}{2D}-\frac{D-1}{D}\alpha=\alpha,
\]
where the equality follows from $(2D-1)\alpha=1/2$. Distributing unassigned goods completes the allocation without decreasing any agent's value.

Finally, the smallest power of two at least $19+\log_2^*n$ is smaller than twice that number. Hence $4\widetilde D(n)-2<8(19+\log_2^*n)$, proving the displayed lower bound.
\end{proof}

\section{A smooth bound for allocating groups of agents}\label{app:envelope}

The family $F_d(x)$ in~\eqref{eq:Fd} records the guarantee from a dyadic number $d$ of groups, where $x=\max\{1,n/q\}$. By increasing $d$, we improve the value reached within each group but incur a larger rounding loss. To obtain the smooth bound $1/(11x-5)$, we compare neighboring choices at their crossing points and use convexity between them.

\begin{lemma}\label{lem:envelope}
For $x\ge1$ and dyadic $d\ge2$, let $F_d(x)=(1-x/d)_+/(2d-1)$ and $x_d=4d^2/(6d-1)$.
If $d$ is the first dyadic integer at least two with $x\le x_d$, then
\[
 F_d(x)\ge\frac{1}{11x-5}.
\]
\end{lemma}

\begin{proof}
We first solve $F_d(x)=F_{2d}(x)$ where both functions are positive. This gives $x=x_d$, with
\[
 F_d(x_d)=F_{2d}(x_d)=\frac{1}{6d-1}.
\]
The inequalities $2d/3<x_d<d<x_{2d}$ show that the crossing points increase without bound, so the specified choice of $d$ exists for every $x\ge1$. We use $d=2$ on $[1,x_2]$ and each $d>2$ on $(x_{d/2},x_d]$. Since $x_d<d$, the function $F_d$ is affine on the corresponding closed interval.

To compare our guarantee with the smooth bound, let $h(x)=1/(11x-5)$. At the initial endpoint, $h(1)=F_2(1)=1/6$. At each crossing point, we have
\[
 (11x_d-5)-(6d-1)=\frac{2(4d-1)(d-2)}{6d-1}\ge0,
\]
so $h(x_d)\le F_d(x_d)=F_{2d}(x_d)$. This also gives the comparison with $F_d$ at $x_{d/2}$ when $d>2$. We have therefore shown that $h\le F_d$ at both endpoints of every selected interval. Since $h$ is convex and $F_d$ is affine there, we conclude that the inequality holds throughout the interval. This proves the bound for every $x\ge1$.
\end{proof}

\begin{remark}
Among the decreasing functions $1/(bx+6-b)$ with $b>0$, all of which pass through $(1,1/6)$, the coefficient $b=11$ is the smallest that gives a lower bound on $\sup_{d\in\{2,4,8,\ldots\}}F_d(x)$ for every $x\ge1$. To see this, we consider the first crossing $x_2=16/11$. Here $F_2(x_2)=F_4(x_2)=1/11$, while $F_d(x_2)\le1/(2d-1)\le1/15$ for all dyadic $d\ge8$. Any such lower bound must therefore satisfy
\[
 \frac{1}{6+5b/11}\le\frac1{11},
\]
and hence $b\ge11$.
The lemma shows that $b=11$ suffices. This optimality concerns only this family of grouping guarantees and lower bounds with the stated affine denominator; it places no upper bound on the MMS approximation obtainable by other methods.
\end{remark}

\section{Bounds for exponential comparisons}
\label{xos:app:exponential}

We prove Lemma~\ref{xos:lem:response} by comparing several exponential
terms with a single term of the same total coefficient. A scaling
argument then gives the required bounds.

\begin{lemma}\label{xos:lem:merge}
Let $Z,Z_1,\ldots,Z_k$ be independent mean-one exponential random
variables. If $b_j\ge0$ and $L=\sum_jb_j\le2$, then
\[
 \Prb\bigl(Z>\max_jb_jZ_j\bigr)\ge\frac1{1+L}.
\]
\end{lemma}
\begin{proof}
The case $L=0$ is immediate. Otherwise, omit zero coefficients and
consider two positive coefficients $a,b$ with $a+b\le2$.
The maximum of the two exponential terms has distribution function
$(1-e^{-t/a})(1-e^{-t/b})$, whereas one competitor with coefficient
$a+b$ has distribution function $1-e^{-t/(a+b)}$.
Let the difference of these distribution functions be
\[
 D(t)=e^{-t/(a+b)}-e^{-t/a}-e^{-t/b}+e^{-t(1/a+1/b)}.
\]
Integrating against the density of $Z$ gives
\begin{equation*}
 \int_0^\infty e^{-t}D(t)\,dt
 =\frac{ab(1-ab)}{(1+a)(1+b)(1+a+b)(ab+a+b)}\ge0,
\end{equation*}
because $ab\le(a+b)^2/4\le1$.

To include the other competitors, we show that $D$ changes sign
once, from negative to positive. Let $\theta=a/(a+b)$, let
$\varepsilon=\theta(1-\theta)$, and let $z=(a+b)t/(ab)$. Then
\begin{equation*}
 e^{(1-\varepsilon)z}D(t)
 =(e^{\theta^2z}-1)(e^{(1-\theta)^2z}-1)-(1-e^{-\varepsilon z}).
\end{equation*}
The right side is strictly convex for $z\ge0$. Indeed, the product
of the nonnegative increasing convex factors is convex, and
$e^{-\varepsilon z}-1$ is strictly convex. The expression vanishes
at zero, has derivative $-\varepsilon<0$ there, and tends to infinity.
It has exactly one positive zero. The same sign change holds for
$D(t)$; denote its positive zero by $t_*$.

Let $F(t)$ be the distribution function of the maximum of the other
competitors, or let $F(t)=1$ when there are none. Monotonicity gives
$(F(t)-F(t_*))D(t)\ge0$ for every $t\ge0$, and hence
\[
 \int_0^\infty e^{-t}F(t)D(t)\,dt
 \ge F(t_*)\int_0^\infty e^{-t}D(t)\,dt\ge0.
\]
Replacing the two coefficients by their sum therefore does not
increase the probability that the unit exponential exceeds all
competitors. Repeat until one competitor remains. Every replacement
is valid because the total is at most two. The final coefficient is
$L$, with comparison probability $1/(1+L)$.
\end{proof}

\begin{proof}[Proof of Lemma~\ref{xos:lem:response}]
We first prove that, for every $0\le\zeta\le1$ and $x\ge0$,
\begin{equation}\label{xos:eq:affine}
 \mathcal R_V(x)\ge
 \frac{1+2\zeta}{(1+\zeta)^2}x
 -\frac1{(1+\zeta)^2}\sum_jb_j.
\end{equation}
The case $x=0$ is immediate. For $x>0$, divide every coefficient
by $x$. Let $b_j$ denote the scaled coefficients, let
$L=\sum_jb_j$, and let $p=\Prb(Z>\max_jb_jZ_j)$.
By~\eqref{xos:eq:memoryless}, it suffices to prove
\begin{equation}\label{xos:eq:unit-affine}
 p\ge\frac{1+2\zeta-L}{(1+\zeta)^2}.
\end{equation}
If $0\le L\le2$, Lemma~\ref{xos:lem:merge} applies, and
\[
 \frac1{1+L}-\frac{1+2\zeta-L}{(1+\zeta)^2}
 =\frac{(L-\zeta)^2}{(1+L)(1+\zeta)^2}\ge0.
\]

Suppose $2\le L\le3$. Scale the competing coefficients to total
two, and let $V$ be the maximum of the corresponding exponential
terms after scaling. Let $X=e^{-V}$
and $t=L/2\ge1$. Lemma~\ref{xos:lem:merge} gives $\E X\ge1/3$.
Jensen's inequality and the tangent inequality for $y\mapsto y^t$
at $y=1$ give
\[
 p=\E X^t\ge(\E X)^t\ge(1/3)^t
 \ge1+t(1/3-1)=1-L/3\ge(3-L)/4.
\]
For $L\ge2$, the right side of~\eqref{xos:eq:unit-affine} is
increasing in $\zeta\in[0,1]$, since its derivative is
$2(L-\zeta)/(1+\zeta)^3\ge0$. Its maximum is $(3-L)/4$, proving
the desired bound on $2\le L\le3$. For $L\ge3$, that right side is
nonpositive. Multiplying by $x$ and using~\eqref{xos:eq:memoryless}
proves~\eqref{xos:eq:affine}.

Taking \(\zeta=1\) in (\ref{xos:eq:affine}) gives the first inequality in (\ref{xos:eq:two-affines}). 
For the second inequality, we require the coefficient of \(\sum_j b_j\) to be at most \(1/2\), which holds when \(\zeta\ge\sqrt2-1\). Since \((1+2\zeta)/(1+\zeta)^2\) decreases with \(\zeta>0\), we choose the nearby rational value \(\zeta=5/12\), for which \(264/289>73/80\) and \(144/289<1/2\). 
This proves the second inequality in (\ref{xos:eq:two-affines}).
For~\eqref{xos:eq:nonlinear}, let $\widetilde v$ be as in the lemma and let
$L=\sum_{g\in\supp(\widetilde v)}b_g\le1$. Summing~\eqref{xos:eq:affine}
on this support with $\zeta=L$, and using $\widetilde v(M)=1$, gives
\[
 \sum_{g\in M}\mathcal R_{V_g}(\widetilde v(g))
 \ge\frac{1+2L-L}{(1+L)^2}=\frac1{1+L},
\]
which proves~\eqref{xos:eq:nonlinear}.
\end{proof}

\section{Additional proofs for XOS valuations}
\label{xos:app:additional}

\subsection{Additive rounding}
\label{xos:app:rounding}

\begin{proof}[Proof of Lemma~\ref{xos:lem:additive-rounding}]
We follow Shmoys and Tardos~\cite{ST93}.
For each agent, order goods by nonincreasing $\widetilde v_i(g)$ and place their
fractional amounts $f_{ig}$ consecutively into slots of capacity one.
There are some full slots and possibly a final partial slot. In the
bipartite graph between full slots and goods, these fractions give a
fractional matching covering every full slot and using each good at
most once. Hall's condition holds: any collection of full slots has
fractional mass equal to its cardinality and is supported on at least
that many goods. An integral matching covering all full slots exists.

Assign each matched good to the owner of its slot. Any good meeting
slot $j$ has value at least every good meeting slot $j+1$, so its value
is at least the fractional value in slot $j+1$, including a partial
slot. Summing over full slots gives the original fractional value
except possibly the first slot, whose value is at most
$\lVert \widetilde v_i\rVert_\infty$. If there is no full slot, the inequality is
immediate.
\end{proof}

\subsection{The final scalar inequality}
\label{xos:app:scalar}

\begin{proof}[Verification of~\eqref{xos:eq:threshold-balance}]
Let $\alpha=5/16$. We verify~\eqref{xos:eq:threshold-balance}
for $0\le\delta\le3/32$ and $0\le s\le3/16-2\delta$. For fixed $\delta$, the left side is
convex in $s$: its only nonlinear term is
\[
 \frac{1/8+(53/100)s}{3/8+\delta+2s}
 =\frac{53}{200}
 +\frac{1/8-(53/200)(3/8+\delta)}{3/8+\delta+2s},
\]
whose last numerator is at least $1/1280>0$. Thus we need only
check $s=0$ and $s=3/16-2\delta$. To express the endpoint
calculations concisely, let $x=16\delta$, so $0\le x\le3/2$.
The right side of~\eqref{xos:eq:threshold-balance} minus the left
side at these endpoints is, respectively,
\[
 \frac{48-26x+6x^2}{75(6+x)(3+x)},
 \qquad
 \frac{82x^3-45x^2-11x+66}
 {300(4-x)(3+x)(6+x)}.
\]
The first numerator is at least $9$. For the second,
$x+x^2\le1+x^3$, since
$(x-1)^2(x+1)\ge0$. Therefore
\[
 82x^3-45x^2-11x+66\ge37x^3+21>0.
\]
Both denominators are positive. This proves
\eqref{xos:eq:threshold-balance}.
\end{proof}

\bibliographystyle{alphaurl}
\bibliography{references}

\end{document}